\documentclass[useAMS,usenatbib]{mn2e}
\usepackage{snapshot}
\usepackage{rotating}
\usepackage{epsfig}
\usepackage{aas_macros}
\usepackage{multirow}
\usepackage{amssymb}
\usepackage{amsmath}
\usepackage{url}

\title[Galaxy Cluster X-ray Luminosity Function]{The WARPS Survey. VIII. Evolution of
  the Galaxy Cluster X-ray Luminosity Function}
\author[L. A. Koens et. al.]{L. A. Koens$^{1,2}$\thanks{E-mail:
lak@roe.ac.uk},
B. J. Maughan$^2$, L. R. Jones$^3$, H. Ebeling$^4$, D. J. Horner$^5$\and
E. S. Perlman$^6$, S. Phillipps$^2$, and C. A. Scharf$^7$\\
\\
$^1$SUPA, Institute for Astonomy, University of Edinburgh, Royal Observatory, Blackford Hill, Edinburgh EH9 3HJ \\
$^2$HH Wills Physics Laboratory, Tyndall Avenue, Bristol, BS8 1TL, UK\\
$^3$School of Physics and Astronomy, The University of Birmingham,
Edgbaston, Birmingham B15 2TT, UK.\\
$^4$Institute for Astronomy, 2680 Woodlawn Drive, Honolulu, HI 96822,
USA.\\
$^5$NASA Goddard Space Flight Center, Code 660.1, Greenbelt, MD 20771, USA\\
$^6$Department of Physics and Space Sciences, Florida Insitute of Technology, 150 West University Boulevard,\\
\hspace{0.1cm}Melbourne, FL 32901, USA \\
$^7$Columbia Astrophysics Laboratory, MC 5247, 550 West 120th St., New York, NY 10027, USA.}

\begin{document}

\date{... Accepted. ... Received; ... in original form; ...}

\pubyear{2012}

\maketitle

\label{firstpage}

\begin{abstract} We present measurements of the galaxy cluster X-ray
  Luminosity Function (XLF) from the Wide Angle ROSAT Pointed Survey
  (WARPS) and quantify its evolution.  WARPS is a serendipitous survey
  of the central region of ROSAT pointed observations and was carried
  out in two phases (WARPS-I and WARPS-II). The results here are
  based on a final sample of 124 clusters, complete above a 
  flux limit of $6.5 \times 10^{-14}$ erg cm$^{-2}$ s$^{-1}$, with members out to 
  redshift $z\sim1.05$, and a sky coverage of 70.9 deg$^2$. We find significant
  evidence for negative evolution of the XLF, which complements the
  majority of X-ray cluster surveys. To quantify the suggested
  evolution, we perform a maximum likelihood analysis and conclude
  that the evolution is driven by a decreasing number density of high
  luminosity clusters with redshift, while the bulk of the cluster
  population remains nearly unchanged out to redshift $z\approx1.1$,
  as expected in a low density Universe. The results are found to be
  insensitive to a variety of sources of systematic uncertainty that
  affect the measurement of the XLF and determination of the survey
  selection function. We perform a Bayesian analysis of the XLF to
  fully account for uncertainties in the local XLF on the measured
  evolution, and find that the detected evolution remains significant
  at the $95\%$ level. We observe a significant excess of clusters in
  the WARPS at $0.1<z<0.3$ and $L_X\approx2\times10^{43}$ erg s$^{-1}$
  compared with the reference low-redshift XLF, or our Bayesian fit to
  the WARPS data. We find that the excess cannot be explained by
  sample variance, or Eddington bias, and is unlikely to be due to
  problems with the survey selection function.
\end{abstract}

\begin{keywords} cosmology: observations - galaxies: clusters: general
- X-rays: galaxies: clusters
\end{keywords}

\section{Introduction} Evolutionary properties of gravitationally
bound objects in the universe are described by models of structure
formation.  The currently favoured cosmology (flat $\Lambda$CDM)
predicts little change in the abundance of galaxy clusters at late
times when the energy density of the universe becomes dominated by
$\Omega_\Lambda$.  The evolution of cluster abundance depends on the
growth rate $f$, which is mainly sensitive to the mean cosmic matter
density $\Omega_m$ as $f(z)\simeq\Omega_m(z)^\gamma$, where
$\gamma\simeq0.6$ in a Universe described by General Relativity \citep{2005PhRvD..72d3529L}.

Galaxy clusters, the largest objects to have decoupled from the Hubble
expansion, are particularly interesting for studying these properties
as a result of their X-ray brightness. The X-ray emission is the
result of bremsstrahlung emitted by the hot intracluster medium ($10^7
- 10^8$K) which contributes more than 80$\%$ of the baryonic content
of the cluster.  Therefore, the mass of a cluster can be estimated from
its luminosity with the use of scaling relations and some simplifying assumptions
\citep{1986MNRAS.222..323K}. The X-ray emitting gas has enabled
cluster detections out to high redshift ($z\gtrsim1$).  Hence, X-ray
galaxy cluster surveys potentially cover a significant portion of the
evolution history of clusters and have high statistical completeness,
thus providing the leverage to place tight cosmological constraints
\citep[e.g.][]{1999ApJ...517...40B,2003A&A...398..867S,vik09a,man10}.

Early predictions of evolution in the number density of clusters,
e.g. \cite{1986MNRAS.222..323K}, pointed towards strong positive
evolution -- an increase in the number density of clusters with
redshift. This prediction assumes a matter power spectrum with a
power-law form, and that the heating of gas is solely by adiabatic
compression during the collapse of dark matter halos. The first
opportunity to test these predictions came with the Einstein Medium
Sensitivity Survey \citep[EMSS][]{gio90a}, which detected clusters out
to $z\approx0.8$. Contrary to the theoretical prediction, the first
teams to test for evolution in the XLF found strong negative evolution
\citep{1990ApJ...356L..35G, edg90, 1992ApJ...386..408H}.  
These controversial findings heated the debate and together with 
the launch of the ROSAT X-ray observatory gave rise to a flurry of 
attempts to measure evolution in the XLF, with some later analyses
raising concerns over the Einstein results \citep[e.g.][]{ell02}.

ROSAT performed an all-sky survey which was used to construct large
flux limited cluster samples, from which the local cluster XLF was
accurately determined.  There are three such surveys: the Brightest
Cluster Sample \citep{ebe98,ebe00}, the ROSAT
All-Sky Survey 1 Brightest Sample \citep{deg99}, and the
ROSAT-ESO Flux-Limited X-ray (REFLEX) galaxy cluster survey
\citep{boh01}. These local XLFs act as the crucial
baseline for quantifying evolution in deeper surveys.

Once the ROSAT all-sky survey was completed the observatory remained
available for pointed observations, which has resulted in an extensive
archive of deep observations, providing the ingredients for many
serendipitous X-ray cluster surveys. This includes the Wide Angle
ROSAT Pointed Survey
\citep{1997ApJ...477...79S,2002ApJS..140..265P,hor08},
the subject of this paper.  Similar surveys that probe the X-ray
universe out to high redshift include the ROSAT International
X-ray/Optical Survey
\citep[RIXOS][]{cas95,2000MNRAS.311..456M}, the ROSAT
Deep Cluster Survey
\citep[RDCS][]{1995ApJ...445L..11R,1998ApJ...492L..21R}, the Bright
Serendipitous High-Redshift Archival ROSAT Cluster (BSHARC) Survey
\citep{2000ApJS..126..209R}, the Massive Cluster Survey (MACS)
\citep{ebe01}, the Brera Multi-scale Wavelet ROSAT HRI
(BMW-HRI) survey \citep{2001ASPC..234..393M,2003A&A...399..351P} ROSAT
North Ecliptic Pole (NEP) Survey
\citep{2001ApJ...553L.109H,gio01}, SSHARC
\citep{bur03}, and the 160 Square Degree
\citep[160SD][]{1998ApJ...502..558V,2003ApJ...594..154M,mul04},
extended to the 400 Square Degree (400SD) survey
\citep{bur07}.

XMM-Newton archival data is also used for surveys based on
serendipitous cluster detections. Currently in progress are the XMM
Cluster Survey \citep[XCS][]{2011arXiv1106.3056M}, the XMM-Newton
Distant Cluster Project \citep[XDCP][]{fas11}, and the
XMM-Newton eXtra eXtra Large (XXL) Survey
\citep{2011MNRAS.414.1732P}.  One serendipitous galaxy cluster survey
is based on Chandra archival data and is part of the Chandra
Multiwavelength Project \citep[CHaMP][]{bar06}.

The most recent determination of the XLF was performed by
\citet{mul04} using 201 clusters from the 160SD catalogue. This work found significant
evidence for negative evolution of the XLF at the bright end. That is,
the number density of high luminosity clusters was lower at
$0.6<z<0.8$ than in the local Universe. Meanwhile \cite{man08} 
used the XLF of several ROSAT cluster surveys at
Ê$z<0.5$ to measure the cluster mass function and hence constrain
Êcosmological parameters.

In this paper we investigate the evolution of the XLF of a sample of
124 WARPS galaxy clusters detected above a flux limit of $6.5 \times
10^{-14}$ erg cm$^{-2}$ s$^{-1}$ over a total area of 70.9 deg$^2$,
and covering a wide redshift range ($0.02 < z < 1.10$).  The survey
design was outlined in \citet{1997ApJ...477...79S} and the catalogues
are presented in two separate papers: WARPS-I
\citep{2002ApJS..140..265P} and WARPS-II
\citep{hor08}. The evolution of the WARPS galaxy
clusters has previously been investigated using phase-I of the
survey. \cite{1998ApJ...495..100J} found no significant evolution in
the log$N$$-$log$S$ relation from the WARPS-I sample and a preliminary
measurement of the XLF (constructed when the survey was complete for $z
< 0.85$) was also found to be consistent with no evolution
\citep{2000lssx.proc...35J}.

This work represents a useful cross-check and extension of the
\citet{mul04} results. While the WARPS survey covers a smaller area,
it is deeper; the 160SD XLF extends to $z\approx0.7$. Importantly,
while both surveys are drawn from ROSAT pointed observations, the
cluster detection and confirmation strategies differ significantly,
allowing us to assess the sensitivity of the evolution results to
those factors.

The current paper is organised as follows: $\S$2 briefly reviews the
WARPS survey and the combined WARPS-I + WARPS-II sample.  The
selection function of the full survey is presented for the first
time. In $\S$3 the X-ray Luminosity function is presented for
different redshift ranges.  Subsequently, in $\S$4 a maximum
likelihood analysis robustly assesses evolution in the XLF. In $\S$5
and $\S$6 we discuss our results and summarise our conclusions.
Throughout the paper errors are quoted at the $68\%$ confidence level
and a $\Lambda$CDM cosmology of $H_0=70\, h_{70}$ km s$^{-1}$ Mpc$^{-1}$, $\Omega_m= 0.3$, and
$\Omega_\Lambda=0.7$, is adopted. All fluxes are corrected for absorption,
and are quoted in the observer's  frame $0.5-2$ keV band.
Luminosities are converted to the rest frame $0.5-2$ keV band of each cluster.

\section{The WARPS Cluster Sample}

\begin{figure}\begin{center}
\includegraphics[scale=0.45]{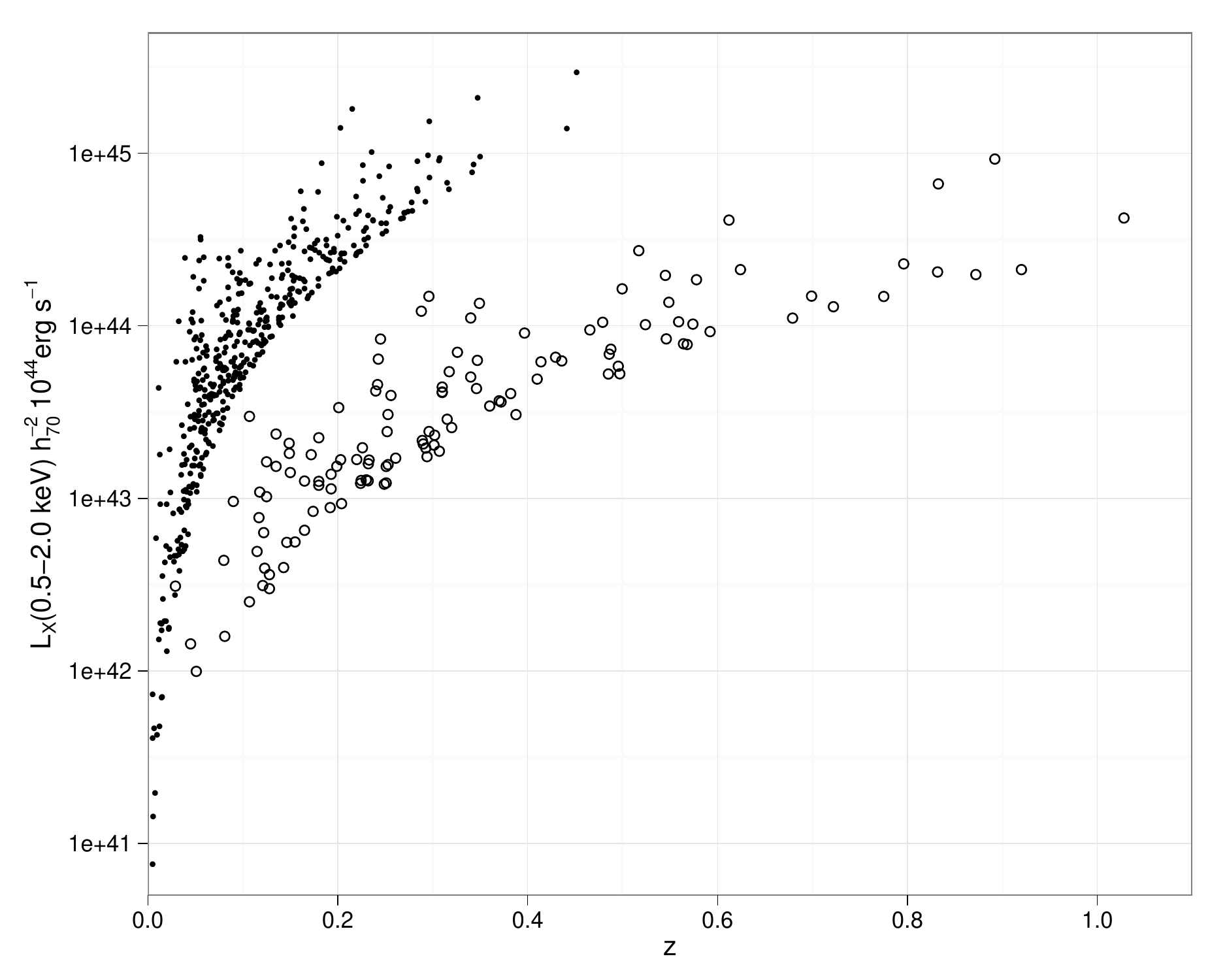}
\caption{The WARPS clusters (hollow points) as points in the $L_\text{X}$-$z$
plane. Also plotted is the REFLEX sample (solid points), which
provides the low-redshift reference XLF for this study \citep{boh01}.}
\label{fig:sample}
\end{center}\end{figure}

The survey is based entirely on serendipitous detections in ROSAT
images from pointed observations with the Position Sensitive
Proportional Counter (PSPC) instrument. Here we summarise the key
facts and direct the reader to \cite{1997ApJ...477...79S} for full
details of the survey methodology.

PSPC fields were selected based on the following criteria. The fields
are at a Galactic latitude of $|b| \geq 20^\circ$, have
exposure times of $t_\text{exp} \geq 8$ks, are non-overlapping, and
the original target is not a galaxy cluster or some other source such
as a bright star that would hamper optical follow-up. Out of the
$\sim$7000 fields in the HEASARC archive 381 satisfy the criteria.

Sources were detected with Voronoi Tessellation and Percolation (VTP)
\citep{ebe93} in an annulus of inner radius $3$
arcminutes and outer radius $15$ arcminutes. VTP does not discriminate
against shape or size and is particularly sensitive to sources of low
surface brightness. WARPS has assessed the efficacy of VTP as a source
detection algorithm by optically imaging all X-ray candidates in
WARPS-I lacking counterparts on existing sky survey plates.

The completeness and efficiency of the VTP detection algorithm were
established with simulations of azimuthally symmetrical clusters,
inserted into PSPC fields. The detected flux is extrapolated to
infinite radius assuming a $\beta$ profile. Although Chandra and
XMM-Newton data have revealed significant substructure in cluster
images up to $z\sim1$, the relatively poor PSPC angular resolution
means that the assumption of spherical symmetry is not expected to
strongly affect the detection efficiency and flux estimation. This
conclusion is supported by the good agreement between the WARPS and
other ROSAT serendipitous surveys that used independent detection
algorithms and selection functions \citep[][; this work]{hor08}. Based
on our simulations a statistically complete sample was defined, comprising 
124 clusters above a conservative flux of $6.5 \times
10^{-14}$ erg cm$^{-2}$ s$^{-1}$ \citep{hor08} (145 sources were confirmed
by WARPS).

For clusters in common, WARPS fluxes were found to be in reasonable agreement with those determined by other serendipitous ROSAT surveys \citep{hor08}.
Spectroscopic redshifts were obtained for all clusters, with 2 or more
concordant redshifts required to confirm a cluster. WARPS did not
obtain near-infrared imaging of cluster candidates, placing an upper
limit on the redshift out to which clusters can be detected. This
limit is $\sim 1.1$, and the uncertainty arising from this is
addressed in section \ref{subsec:zmax}.

In combining WARPS-I and WARPS-II catalogues, it was found that
background levels were missing for 1 WARPS-I field and 27 WARPS-II
fields. The background level of each field is required in order to
compute the selection function, and so these were remeasured from the
archived PSPC data. The ROSAT PSPC data have been reprocessed since
the cluster detection was performed, so we checked the background
measurements for all WARPS-I fields using the currently available PSPC
data against our original measurements. The new measurements were
found to be $\sim7\%$ lower on average, depending somewhat on the
source detection algorithm used to exclude sources in each field. We
thus renormalised the background measurements for the 28 missing
fields in the combined WARPS catalogue by this factor, for consistency
with the data used for cluster detection. We investigated the impact
of this systematic effect on the selection function, and found it to
be insensitive to whether or not this background scaling was applied
to the 28 missing fields. This is not surprising given the small
magnitude of the correction and the small fraction of fields affected.

In figure \ref{fig:sample} the WARPS clusters are plotted in the
luminosity-redshift plane. The fluxes of the clusters have been K-corrected 
to the cluster rest frame assuming an APEC thermal plasma model 
\citep{2001ApJ...556L..91S}, for which we
set the metallicity to 0.3 $Z_\odot$. The plasma temperature required
for this conversion was estimated iteratively from the X-ray
luminosity using the luminosity temperature scaling relation of
\cite{mar98a}, although the magnitude of the K-correction
was insensitive to this choice.

\begin{figure}\begin{center}
\includegraphics[scale=0.5]{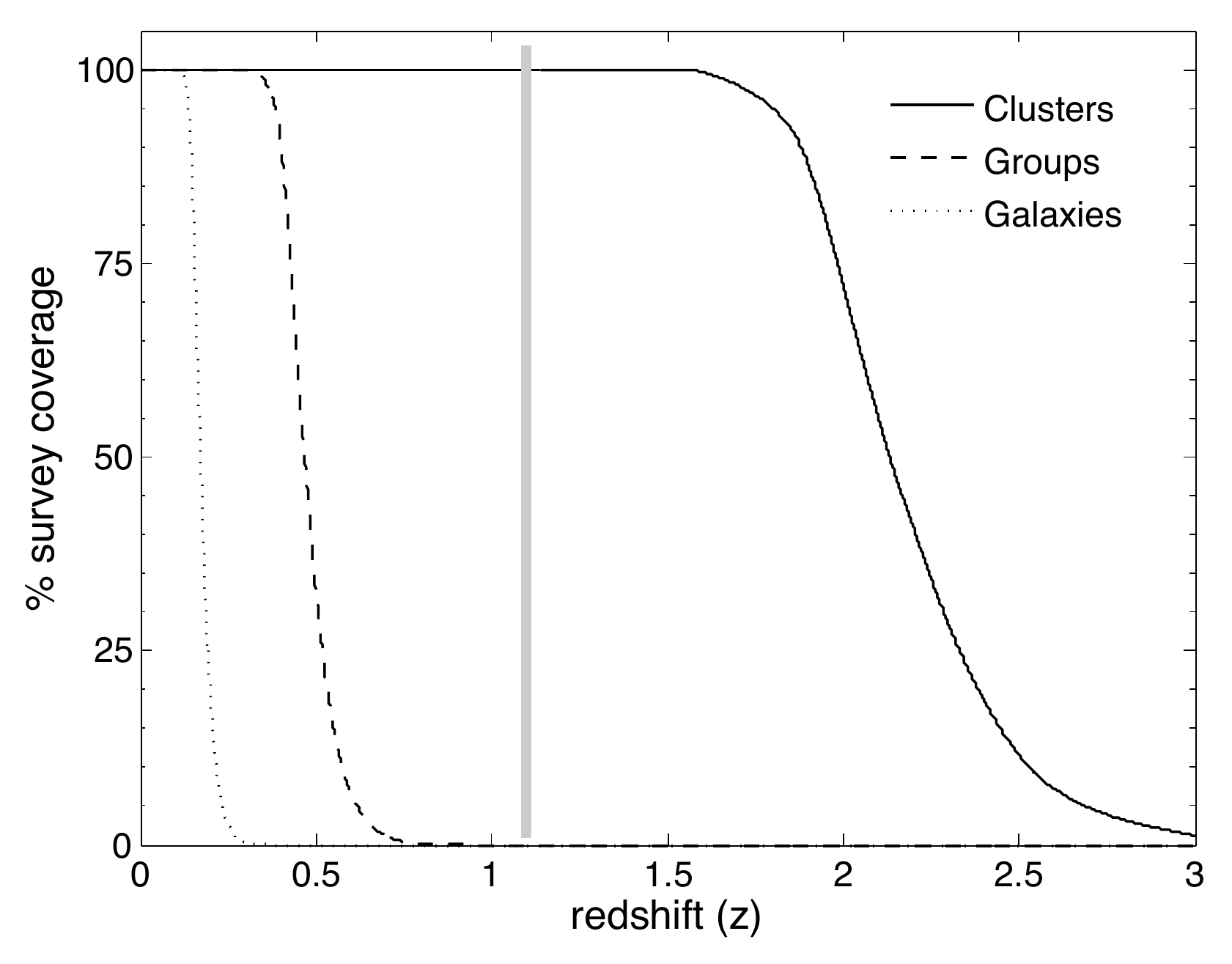}
\caption{Survey coverage for three classes of objects as a function of
redshift. The objects are defined as: elliptical galaxies with
$L_\text{X}(0.5-2.0$ keV$)=1\times10^{42}$ erg s$^{-1}$ and effective
core radius $r_c = 50$ kpc, groups with $L_\text{X}(0.5-2.0$
keV$)=1\times10^{43}$ erg s$^{-1}$ and effective core radius $r_c =
100$ kpc, and clusters with $L_\text{X}(0.5-2.0$ keV$)=5\times10^{44}$
erg s$^{-1}$ and effective core radius $r_c = 250$ kpc. The grey line
represents the approximate upper redshift limit imposed by the lack of
near infra-red follow-up of cluster candidates.}
\label{fig:selfun}
\end{center}\end{figure}

The selection function for WARPS-I and WARPS-II combined is shown in
figure \ref{fig:selfun} and is based on 381 PSPC fields. The effective
sky coverage for an object of given luminosity and extent is
determined by the performance of VTP, the degrading PSF with off-axis
angle, and the background levels and exposure times of the fields. The
variance in the field properties of the survey alters the steepness of
the decrease; e.g. if all the fields were the same, we expect a much
more sudden drop from 100$\%$ to 0$\%$. We find the curve for the full
survey to be very similar to that for WARPS-I, figure 9 in
\citet{1997ApJ...477...79S}.

\section{The X-ray Luminosity Function}
\label{sec:XLF} The X-ray Luminosity Function (XLF), conventionally
given the symbol $\phi$, is the comoving number density $n$ of objects per luminosity
interval:
\begin{equation} \phi(L_\text{X},z) =
\frac{\mathrm{d}n(L_\text{X},z)}{\mathrm{d}L_\text{X}}.
\label{eq:XLFdef}
\end{equation} The Schechter function \citep{1976ApJ...203..297S} is
the canonical, parametric representation of the luminosity function:
\begin{equation} \phi(L_\text{X},z)\mathrm{d}L_\text{X} =
\phi^*\left(\frac{L_\text{X}}{L^*_\text{X}}\right)^{-\alpha}
\text{exp}\left(-\frac{L_\text{X}}{L^*_\text{X}}\right)\left(\frac{\mathrm{d}L_\text{X}}{L^*_\text{X}}\right),
\end{equation} where the parameter $\phi^*$ normalises the XLF, and
$\alpha$ determines the steepness at $L_\text{X} < L_\text{X}^*$.

The conventional method to compute the differential XLF is the $1/V_\text{max}$ 
method  \citep{1968ApJ...151..393S,ab80}, where $V_\text{max}$ 
denotes the maximum co-moving volume, given by
\begin{equation} V_\text{max} =
\int_{z_\text{min}}^{z_\text{max}}\Omega(f_\text{X},r_\theta)\frac{\mathrm{d}V(z)}{\mathrm{d}z}\mathrm{d}z,
\end{equation} 
where $\Omega(f_\text{X},r_\theta)$ is the sky coverage
as a function of flux $f_\text{X}(L_\text{X},z)$ and angular extent
$r_\theta(r_c,z)$ (here $r_c$ is the core radius of the cluster
surface brightness distribution, conventionally parameterised with a
$\beta$-model), and $\mathrm{d}V(z)/\mathrm{d}z$ is the
differential, co-moving volume, which is strongly sensitive to the
cosmological framework. The maximum co-moving volume is calculated for
all $N$ galaxy clusters. The XLF is then obtained by summing the
corresponding density contributions per luminosity bin, that is
\begin{equation} \phi(L_{\text{X}_j},z) = \frac{1}{\Delta
L_{\text{X}_j}}\sum_{i=0}^{N_j}\frac{1}{V_{\text{max},i}},
\end{equation} where the subscript $j$ denotes the $j$-th bin. Due to
the sensitivity to the choice of binning, the method is less ideal for
quantifying evolution. However, it is a conventional way of presenting
a sample of objects, so we include it here to allow easy comparisons with previous
work.

Alternatively, \cite{2000MNRAS.311..433P} provide an estimate of
$\phi$, which expression is obtained by integrating \eqref{eq:XLFdef}
and noting that $\phi$ changes little compared to the survey volume
element in the volume - luminosity plane, such that it can be taken
out of the integral, giving
\begin{equation} \phi(L_{\text{X}_j},z) =
\frac{N_j}{\int_{L_\text{X,min}}^{L_\text{X,max}}\int_{z_\text{min}}^{z_\text{max}}\Omega(f_\text{X},r_\theta)\frac{\mathrm{d}V(z)}{\mathrm{d}z}\mathrm{d}z\mathrm{d}L_\text{X}},
\label{eq:PC}
\end{equation} where $L_{\text{X}_j}$ is the bin centre and $N_j$ is
the number of clusters in the $j$-th bin.

We apply the method of \cite{2000MNRAS.311..433P} to account for the
flux limit of the survey to effectively decrease the width of some of
the bins, enhancing the XLF. The Page-Carrera estimator was also
deployed by \cite{mul04}, who found a marginal increase
at the faint end of the XLF compared to the $V_\text{max}$
estimator. Our results were similarly insensitive to the choice of volume estimator; the uncertainties at the faint end
of the XLF are dominated by those arising from small number statistics, the statistical error on $f_X$ and the uncertainty on $r_\theta$.

\subsection{The WARPS XLF}

In order to present the binned WARPS XLF, we divide the clusters
according to their redshifts to study the local $(0.02 < z < 0.3; 67$ clusters),
intermediate redshift $(0.3 < z < 0.6; 44$ clusters), and high redshift $(0.6 < z <
1.1; 13$ clusters) populations, similar to \cite{mul04}.

We apply the same
$L_\text{X}$ binning as \cite{mul04} to allow for comparison. Poisson
errors on the counts in each luminosity bin are provided by
\cite{geh86}, which are much larger than the flux
measurement errors.

\begin{figure}\begin{center}
\includegraphics[scale=0.44]{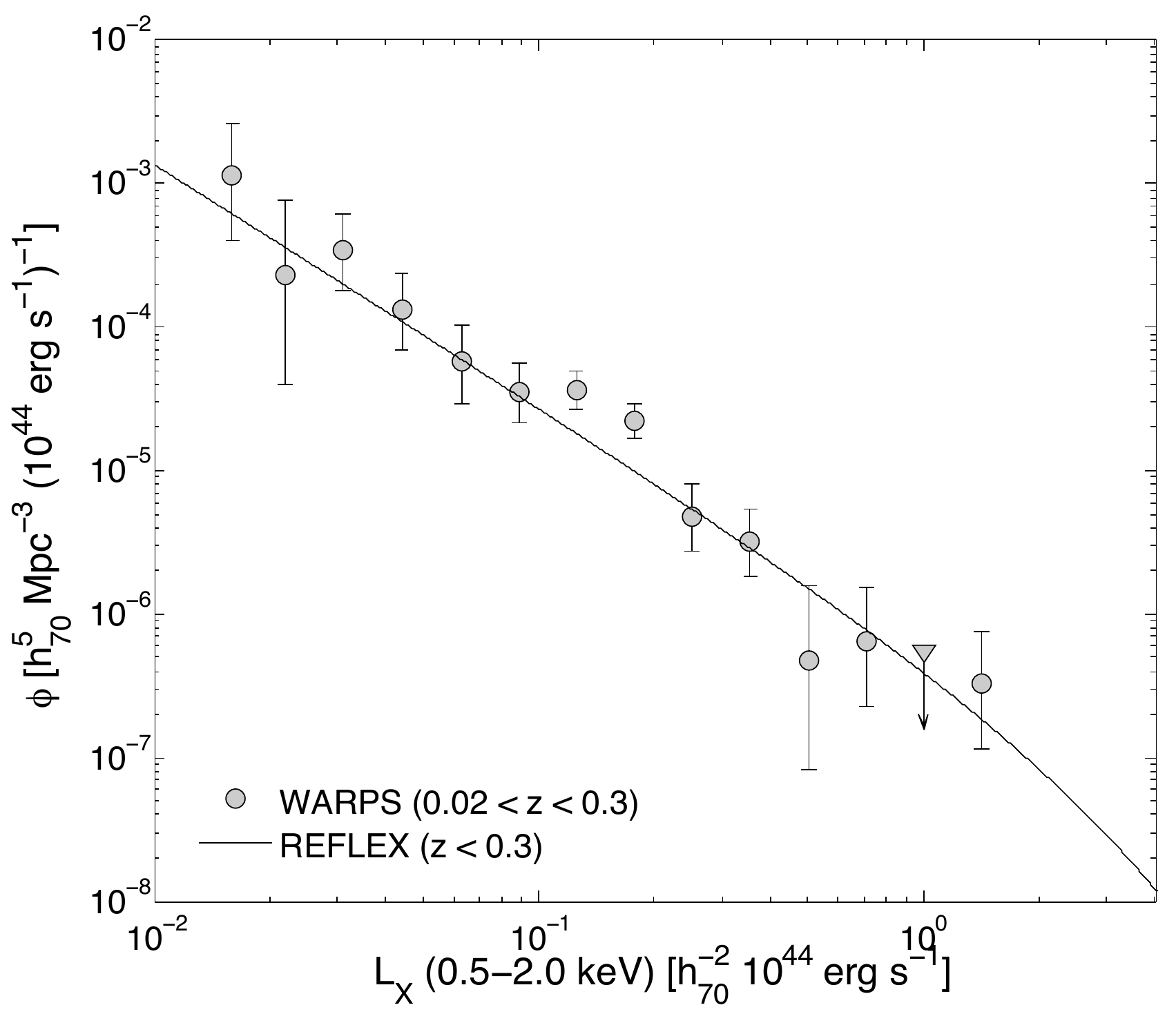}
\caption{The XLF from the local WARPS sample along with the best fit
Schechter function of the REFLEX sample.}
\label{fig:XLFlocal}
\end{center}\end{figure}

Good knowledge of the local XLF is essential for studying its
evolution and is provided with great accuracy by the ROSAT all-sky
survey. The XLF of the local WARPS sample of 67 $0.02<z<0.3$ clusters
is shown in figure \ref{fig:XLFlocal}. The lower redshift limit is set
to $z=0.02$ below which many clusters become too extended relative to
the size of the PSPC fields to be detected. Over this redshift range
the WARPS XLF agrees remarkably well with the all-sky samples,
represented by the REFLEX model in figure \ref{fig:XLFlocal}.

There appears to be a high number density of clusters at
$L_{\text{X}(0.5-2.0\text{keV})} \approx 1.5 \times 10^{43}$ erg
s$^{-1}$ compared to the Schechter function. We note that this feature
is also present in the local XLF of the 160SD sample \citep[figure
4]{mul04}
We test the significance of this excess in section
\ref{subsec:numbers}, and discuss possible interpretations in section
\ref{sec:excess}.

\begin{figure}\begin{center}
\includegraphics[scale=0.42]{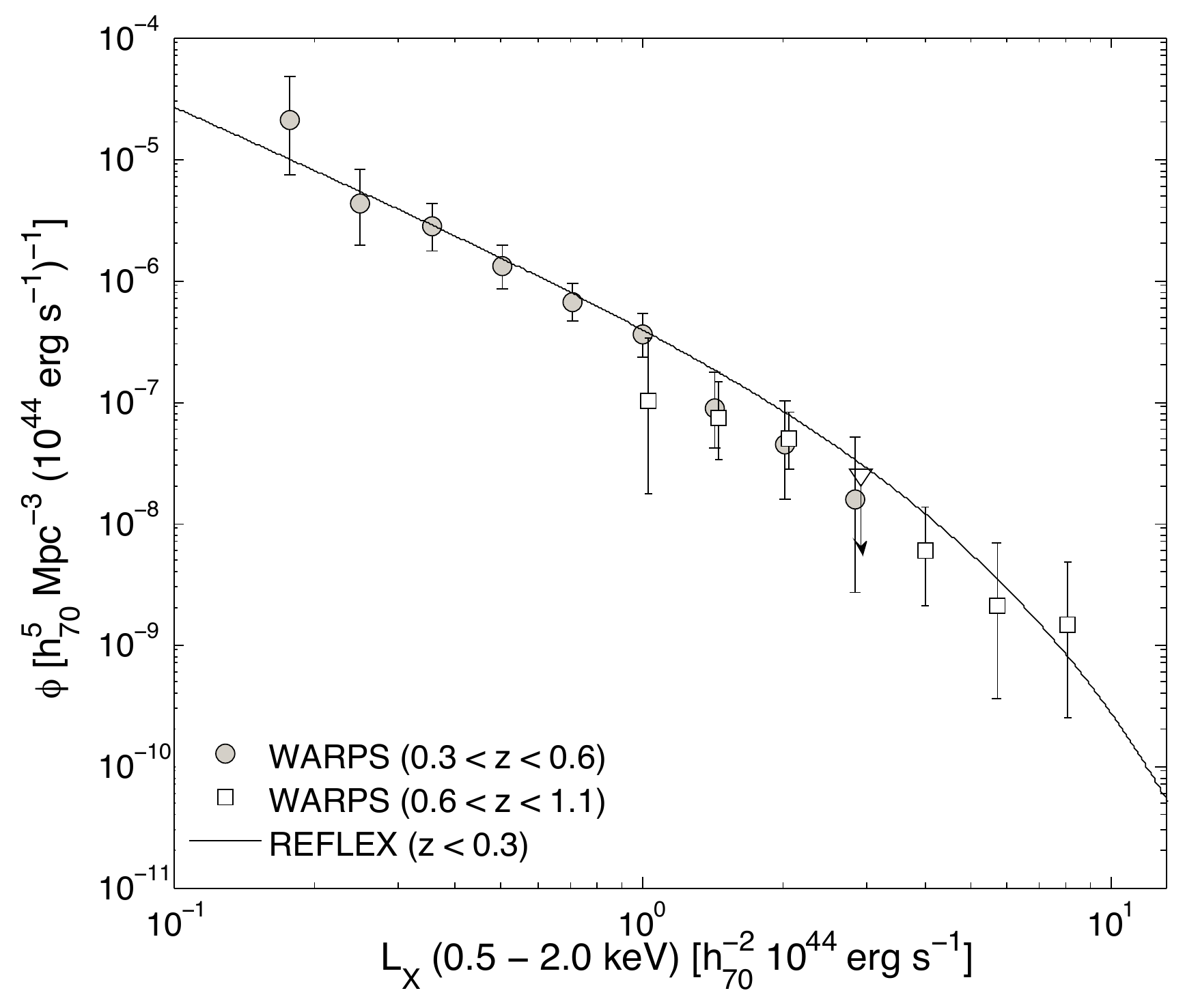}
\caption{The XLF from the intermediate and high redshift WARPS samples
along with the best fit Schechter function of the REFLEX sample.}
\label{fig:XLFhigh}
\end{center}\end{figure}

In figure \ref{fig:XLFhigh} we show the intermediate and high redshift
XLFs along with the local REFLEX Schechter function. The majority of
data points of both the intermediate and high redshift XLF are
slightly low compared to the local baseline. This is a first
indication from the data of negative evolution. Whether
this is significant will be addressed in the next section.

\subsection{Expected Versus Observed Numbers}
\label{subsec:numbers} The expected number of objects in the
luminosity-redshift plane is obtained by integrating equation
\eqref{eq:XLFdef}

\begin{equation}
  N_\text{exp} = \int_{L_\text{X,min}}^{L_\text{X,max}}
  \int_{z_\text{min}}^{z_\text{max}} \phi(L_\text{X},z)
  \Omega(f_\text{X},r_\theta) \frac{\mathrm{d}V(z)}{\mathrm{d}z}
  \mathrm{d}z\mathrm{d}L_\text{X}.
\label{eq:Nexp}
\end{equation}
As mentioned in section \ref{sec:XLF} the XLF changes little compared
to the volume element. Hence we can predict the number of clusters
for any of the WARPS subsets based on the local reference
XLF $\phi_\text{local}$, the observed XLF for the subset
$\phi_\text{observed}$, and the number of clusters observed in that
subset $N_\text{observed}$:
\begin{equation}
  N_\text{exp} \approx N_\text{observed} \times
  \frac{\phi_\text{local}}{\phi_\text{observed}}.
\label{eq:Nexpapprox}
\end{equation}
If the local reference XLF is a good description of the WARPS XLF, and
there is no evolution, then $N_\text{exp}$ should be consistent with
$N_\text{observed}$ for all subsets.

\begin{figure}
\begin{center}
\includegraphics[scale=0.4]{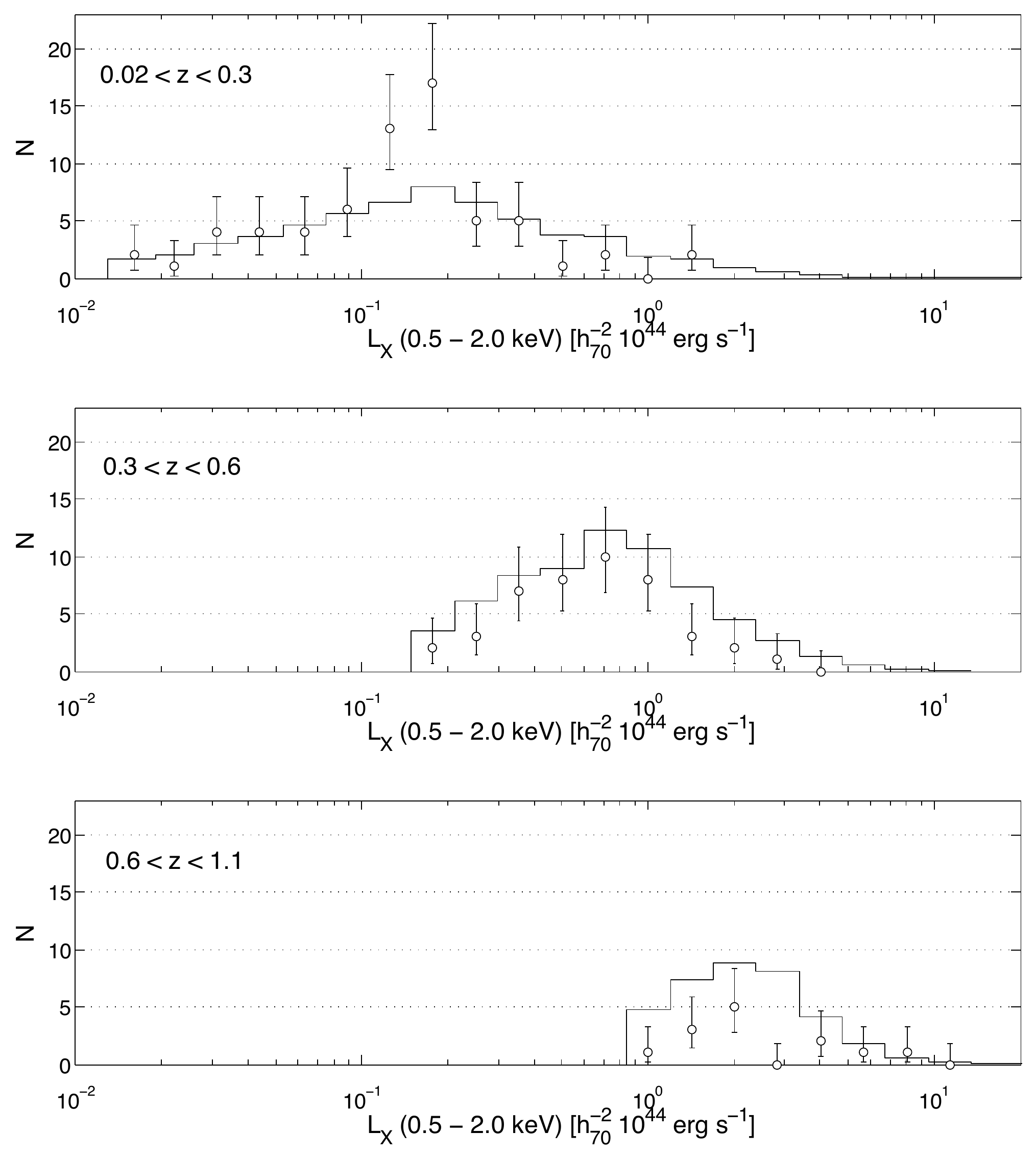}
\caption{Expected cluster numbers (solid line) versus observed (dots)
per luminosity bin for the local, intermediate, and high redshift  samples.
The expected number of clusters is calculated per luminosity bin from
equation \eqref{eq:Nexp}.}
\label{fig:Nexpected}
\end{center}
\end{figure}

Using the REFLEX best fit Schechter function as the local reference,
we compute the expected cluster numbers for each luminosity bin in
each of the WARPS subsets. The results are plotted in figure
\ref{fig:Nexpected}. When integrated over the full range of
luminosities, 60 clusters are expected from equation
\eqref{eq:Nexpapprox} for the low-z subset, instead of the 67
observed. For the intermediate-z subset, 67 are predicted instead of
the 44 observed, and for the high redshift subset, the local relation
predicts 36 clusters instead of the 13 that are observed. The
differences for the low and intermediate redshift subsets are
not strongly significant, but the lack of high-z clusters compared to
the local prediction is significant at $>4\sigma$, assuming Poisson
errors on both numbers.

Figure \ref{fig:Nexpected} also illustrates the excess of clusters
around $L_{\text{X}(0.5-2.0\text{keV})} \approx 1.5 \times 10^{43}$
erg s$^{-1}$ in the low-z subset seen in figure
\ref{fig:XLFlocal}. Over the two bins with excess counts, there are 28
clusters observed, while only 14 are predicted by the REFLEX XLF. 
This is a significant excess; the probability of observing $N>27$ for a
Poisson distribution with a mean of 14 is $6.4\times10^{-4}$. The same
analysis was also applied to the larger low-z subset of the 160SD
sample. According to equation \eqref{eq:Nexpapprox}, the number of
clusters predicted by the local REFLEX XLF over the two bins with
excess counts for the 160SD sample is $18$, significantly lower than
the observed number of $40$ clusters [$P(N>39)=5.3\times10^{-6}$ for a
Poisson distribution with mean $18$].

 The number of clusters, $N_{mod}$, predicted by the REFLEX fit is
  uncertain due to the errors on the REFLEX Schecter function
  parameters. Assuming that the covariance in the REFLEX Schechter
  function parameters is similar to that found in our fit (section
  4.1.3), then the resulting uncertainty on the REFLEX $N_{mod}$ is
  estimated to be $26\%$.  As the observed number counts are a
  Poissonian realisation of the model prediction, the probability of
  observing $\geqslant N_\text{obs}$ clusters for a set of model
  parameters ${\bf \theta}$ which predicts a number $N_\text{mod}$ is
  \begin{align}
  P(N_\text{obs}|{\bf \theta}) & = \int P(\geqslant
  N_{obs}|N_\text{mod}) P(N_\text{mod}|{\bf \theta}) dN_\text{mod}.
  \end{align}
  We model the first probability distribution as a Poisson distribution
  and the second as a Gaussian with mean 14 and standard deviation 3.7,
  which results in a probability of observing at least 28 clusters in
  this luminosity bump of $1\%$. The corresponding probability for the
  bump in the 160SD sample is $0.1\%$.

\subsection{Evolving Schechter function}
\label{subsec:ML} Here we deploy the maximum likelihood analysis first
set out by \cite{1983ApJ...269...35M}, which fits an evolving
Schechter function to the distribution of objects in luminosity
redshift space. The treatment is free from arbitrary binning and with
the generalisation of \cite{mul04} accounts for flux
uncertainties. We briefly summarise the method and apply it to the
WARPS sample.

The XLF is characterised as an evolving Schechter function
\begin{equation}
  \phi(L_\text{X},z)\mathrm{d}L_\text{X}
  =\phi^*(z)\left[\frac{L_\text{X}}{L^*_\text{X}(z)}\right]^{-\alpha}
  \text{exp}\left[-\frac{L_\text{X}}{L^*_\text{X}(z)}\right]\left[\frac{\mathrm{d}L_\text{X}}{L^*_\text{X}(z)}\right]
\end{equation}
The parameters, except for $\alpha$, are allowed to
evolve as follows
\begin{align}
  \phi^*(z) &= \phi_0^*
  \left[\frac{1+z}{1+z_0(L_\text{X})}\right]^A, \label{eq:evA}\\
  L_\text{X}^*(z) &= L_\text{X,0}^* \left[\frac{1+z}{1+z_0(L_\text{X})}\right]^B, \label{eq:evB}
\end{align}
where $\phi_0^*$ and $L_\text{X,0}^*$ are adopted from the
local XLF. Due to the flux limit of the surveys, the median redshift
$z_0$ increases with luminosity bin and is given by the local XLF. A
deviation from $A=B=0$ indicates evolution.

To be free from arbitrary binning, the luminosity redshift grid is
chosen to be sufficiently fine for there to be either 1 or 0 clusters
in each cell. We achieve this with $\mathrm{d}z=0.01$ and
$\mathrm{d}L_\text{X}=0.1 \times 10^{43} h^{-2} $ erg s$^{-1}$ for the
WARPS sample. In each cell the expected number of clusters is
calculated
\begin{equation}
  \lambda(L_\text{X},z)\mathrm{d}L_\text{X}\mathrm{d}z =
  \phi(L_\text{X},z)\Omega(f_\text{X},r_\theta)
  \frac{\mathrm{d}V(z)}{\mathrm{d}z} \mathrm{d}L_\text{X}\mathrm{d}z
\end{equation}
The likelihood function $\mathcal{L}$ describes the
joint probability of detecting 1 cluster at each occupied cell $i$ and
0 in each empty cell $j$ and is given by
\begin{align}\label{eq:likelihood}
\mathcal{L} = \prod_i
&\lambda(L_\text{X,i},z_i)\mathrm{d}L_\text{X}\mathrm{d}z \,\,
\text{exp}\left[-\lambda(L_\text{X,i},z_i)\mathrm{d}L_\text{X}\mathrm{d}z\right]
\notag\\ &\times \prod_j
\text{exp}\left[-\lambda(L_\text{X,i},z_i)\mathrm{d}L_\text{X}\mathrm{d}z\right],
\end{align}
which makes use of the Poisson distribution and is valid
when the number of expected clusters $\ll1$ as expected for small
cells. To account for the uncertainties on the measured fluxes, we smooth the objects by a Gaussian in the luminosity
direction, in the same way as in \cite{mul04}. The amount of smoothing is based on the $1\sigma$ flux errors. Redshift errors are
not taken into account, since they are typically much smaller.

\begin{figure}\begin{center}
\includegraphics[scale=0.42]{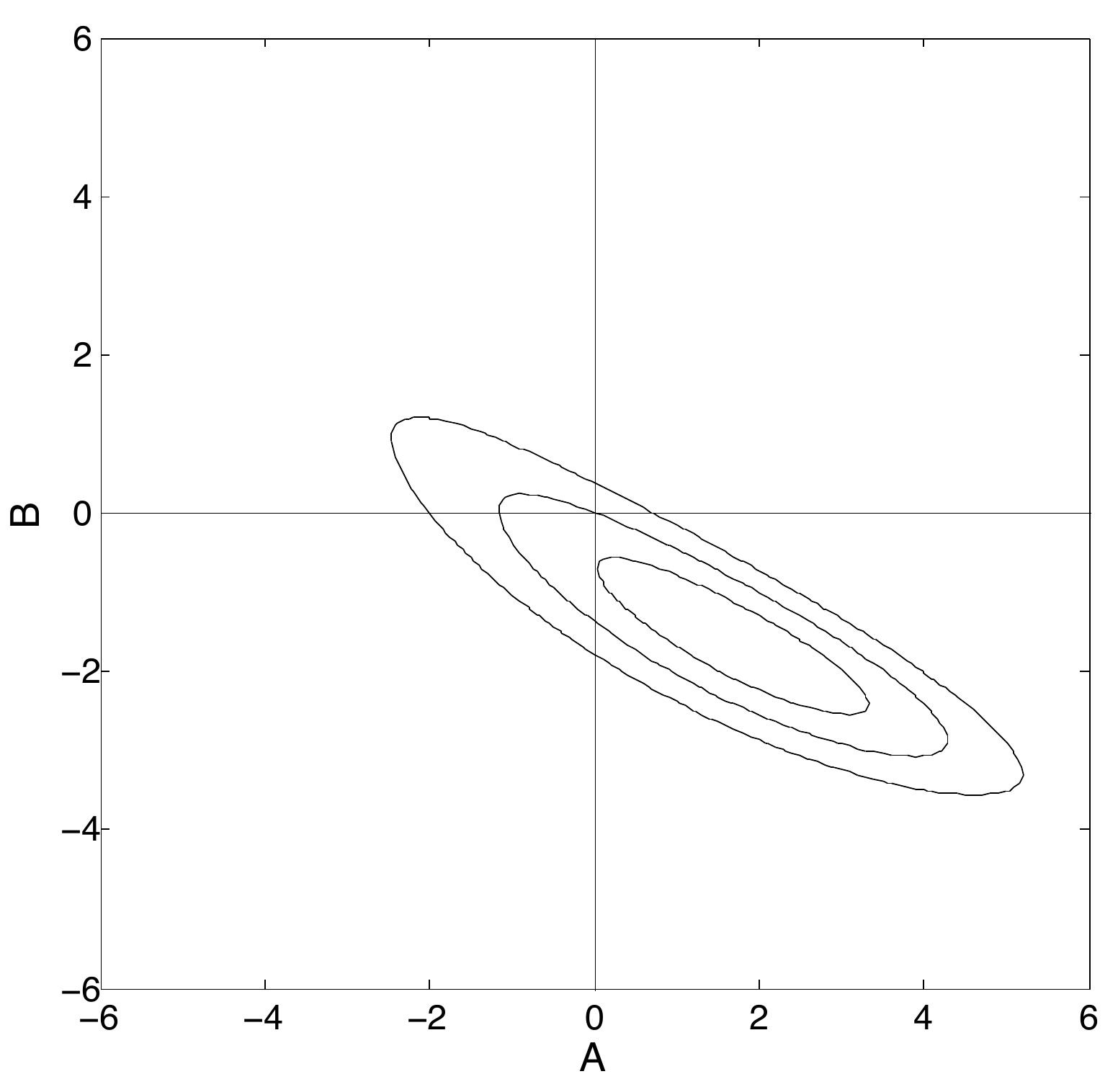}
\caption{Likelihood contours for the evolution parameters $A$ and $B$,
defined in equations \eqref{eq:evA} and \eqref{eq:evB},
based on a comparison of the local REFLEX Schechter function and the
complete WARPS distribution of clusters in luminosity redshift
space. Contours show the $1\sigma$, $2\sigma$, and $3\sigma$
confidence limits.}
\label{fig:ABall}
\end{center}\end{figure}

We calculate $\Delta S = S(A,B) - S(A_\text{best},B_\text{best})$,
where $S = -2 \, \text{ln} \, \mathcal{L}$.  In figure \ref{fig:ABall}
we plot contours of $\Delta S = 2.30$, $6.17$, and $11.8$, which correspond to the
$1\sigma$, $2\sigma$, and $3\sigma$ confidence limits.

\begin{figure}\begin{center}
\includegraphics[scale=0.42]{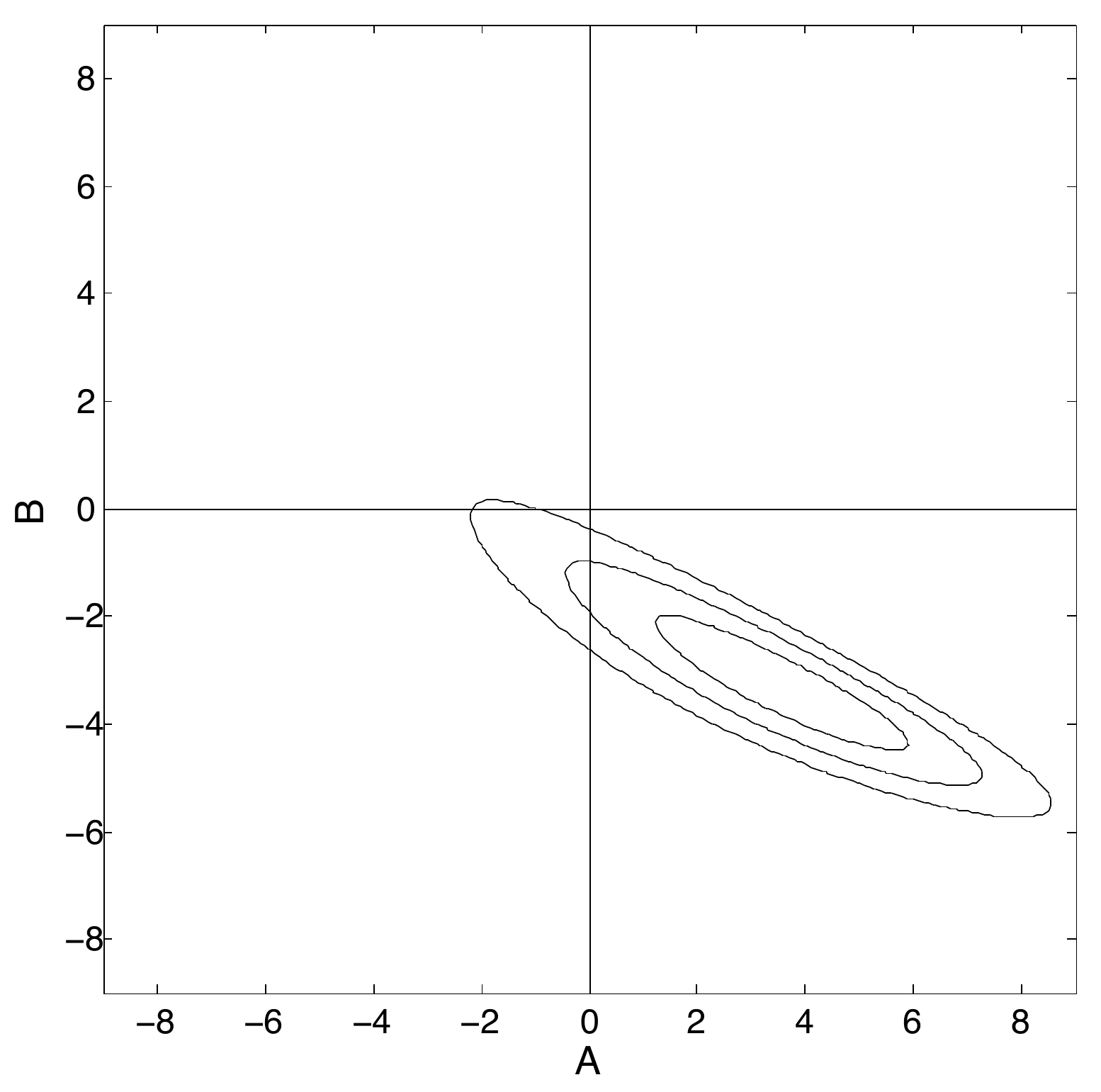}
\caption{Likelihood contours for the evolution parameters $A$ and $B$, 
defined in equations \eqref{eq:evA} and \eqref{eq:evB},
based on a comparison of the local REFLEX Schechter function and the
$z>0.3$ WARPS clusters. Contours show the $1\sigma$, $2\sigma$, and $3\sigma$
confidence limits.}
\label{fig:ABzgt3}
\end{center}\end{figure}

The contours for WARPS are shown in figure \ref{fig:ABall}, with
evolution measured relative to the local XLF from REFLEX, for which we
used all 124 WARPS clusters. We find evidence for negative evolution
that is significant at $2\sigma$, with $A=-1.88\pm0.62$ and
$B=-1.76\pm0.53$. Stronger evidence for evolution was measured 
using only the WARPS clusters at $z>0.3$. The same methodology was 
applied, and the resulting confidence contours on $A$ and $B$ are 
shown in figure \ref{fig:ABzgt3}. The best fitting parameters for both 
samples are given in table \ref{tab:xlf}.

\begin{table*}
\begin{tabular}{llccccc}
\hline
Dataset & redshift & $\phi^*$ & $\alpha$ & $L_X^*$ & $A$ & $B$ \\
 & & $10^{-7}h_{70}^3$Mpc$^{-3}$ & &  $10^{44}h_{70}^{-2}$ & \\
\hline
REFLEX & $z<0.3$ & $2.94\pm0.82$ & $1.690\pm0.045$ & $2.64\pm0.29$ & - & - \\
WARPS ML & $z>0.02$ & $2.94$ & 1.690 & 2.64 & $1.88\pm0.62$ & $-1.76\pm0.53$ \\
WARPS ML & $z>0.3$ & $2.94$ & 1.690 & 2.64 & $3.60\pm0.95$ &
$-3.37\pm0.56$ \\
WARPS Bayesian & $z>0.02$ & $3.68\pm0.87$ & $1.79\pm0.04$ & $2.59\pm0.35$ &
$-0.09\pm1.19$ & $-0.93\pm0.58$ \\
\hline
\end{tabular}
\caption{Best fitting XLF parameters. The REFLEX parameters are taken
  from \citep{boh02}, for a $\Lambda$CDM cosmology. The maximum
  likelihood (ML) fits assumed XLF shape parameters fixed at the REFLEX best
fit values. The Bayesian fit used the REFLEX values as priors, as
discussed in section \ref{sec:bayes}.
\label{tab:xlf}}
\end{table*}

\section{Discussion}
\label{sec:discussion}

\subsection{Evolution in the XLF}
\begin{figure}\begin{center}
\includegraphics[scale=0.42]{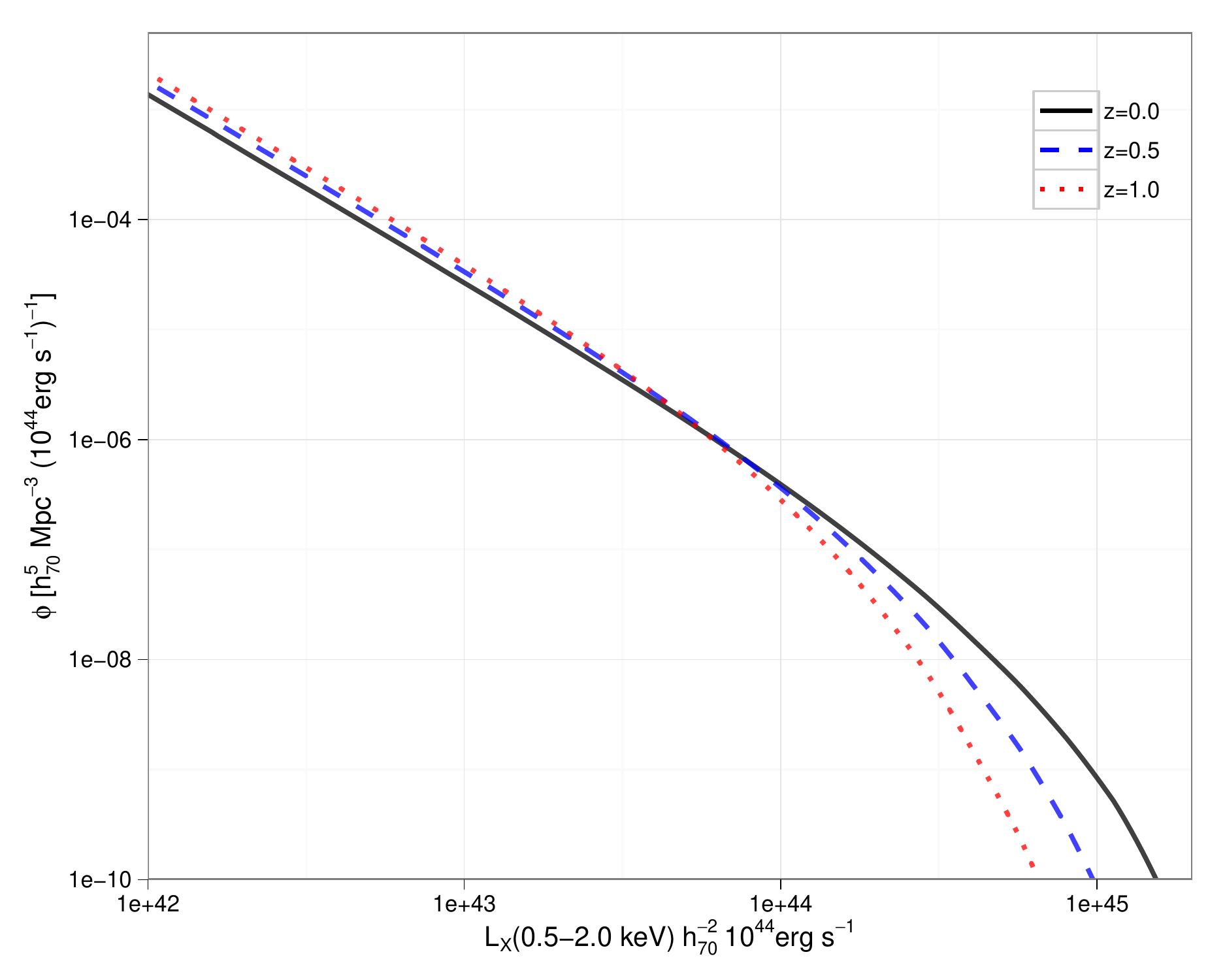}
\caption{Schechter functions with best-fitting evolution parameters
  from the maximum likelihood analysis to the full sample. The
  lines show the form of the Schechter function at redshifts 0, 0.5 and
  1.}
\label{fig:schecter}
\end{center}\end{figure}

The comparison of expected and observed cluster number counts, and
the maximum likelihood analysis of the unbinned cluster population
both strongly support negative evolution of the XLF. The evolution in
$\phi^*$ and $L_X^*$ is degenerate, as is apparent in figure
\ref{fig:ABall}, but the net effect is significant, and consistent
with a decrease in the number density of massive, high luminosity
clusters with redshift, as expected in a $\Lambda$CDM hierarchical
Universe. This is illustrated in figure \ref{fig:schecter}, which
shows the best fitting evolution of the REFLEX $z=0$ Schechter
function.

In the maximum likelihood analysis of section \ref{subsec:ML} we
assume no evolution in the parameter $\alpha$. Here we test whether
this is justified by calculating $\Delta S$ for the parameter C
defined as
\begin{equation}
\alpha(z) = \alpha_0
\left[\frac{1+z}{1+z_0(L_\text{X})}\right]^C
\end{equation}
and use our best fit parameters $A$ and $B$ from
section \ref{subsec:ML}. We find $C_\text{best} = 0.05\pm0.17$, and
for $A=B=0$ we have $C_\text{best} = -0.05\pm0.17$. This is
reassuring, since in the hierarchical picture of structure formation
we expect evolution to occur at the bright end, whereas $\alpha$
determines the slope at fainter luminosities.

The negative evolution measured in the WARPS XLF is in very good
agreement with that found for the 160SD (particularly for the $z>0.3$
subset), and agrees qualitatively with previous measurements of
negative evolution in the XLF \citep[see figure 8 in][]{mul04}. We
note that the 160SD evolution was measured with respect to the BCS
XLF, whereas we used the REFLEX XLF for our low-redshift baseline, so
the agreement in evolution measures suggests that the choice of
baseline does not strongly affect the measured evolution. The
uncertainty on the local XLF and other systematics affecting the
evolution measurements are discussed below.

\subsubsection{Uncertainty on $z_\text{max}$}
\label{subsec:zmax}
The upper redshift limit $z_\text{max}$ for the high redshift sample
is uncertain due to lack of near infra-red follow-up of cluster
candidates. To our knowledge it is the first time this issue has been 
considered in the determination of
the cluster XLF. The RDCS survey included near infra-red
imaging which resulted in the successful detection of 4 clusters
beyond redshift 1. However, these fall below the flux limit of their
statistically complete sample, which has $z_\text{max} = 0.83$. The
detection and optical confirmation in WARPS of a cluster at $z=1.028$
is consistent with our estimate of $z_\text{max}\approx1.1$, but the
exact limit will depend on the characteristics of the galaxy
populations and the photometric limits of the optical imaging for each
cluster, so is not well defined.

As can be seen in figure \ref{fig:selfun}, for very luminous clusters
the sky coverage of 100$\%$ is maintained well beyond the limit set by
the optical observations. This means that the product
$\Omega(f_\text{X},r_\theta) \times
\frac{\mathrm{d}V(z)}{\mathrm{d}z}$ is nonzero in equation
\eqref{eq:PC}.  Hence, an increase in $z_\text{max}$ suppresses the
XLF in those bins that represent sufficiently high X-ray
luminosities. Although the influence of the choice of $z_\text{max}$
on the XLF is suppressed by the flux limit, a too high value for
$z_\text{max}$ could falsely suggest negative evolution, whereas a too
low value boosts the bright end towards the opposite conclusion.

We tested the robustness of the evolution measurement to $z_{max}$ by
reducing $z_{max}$ to $z=1$ and excluding the $z=1.028$ cluster from
the analysis. The contours in the $A-B$ plane were changed
negligibly. This is a conservative approach, and shows our measured
evolution is insensitive to the choice of $z_{max}$. If the effective
$z_{max}$ of the survey is actually larger than the assumed $z=1.1$,
then our non-detection of clusters beyond $z=1.03$ would imply
stronger evolution than that measured here.

\subsubsection{Cluster Surface Brightness}
Although we refer to the statistically complete WARPS sample as
``flux-limited'', in practice it is the X-ray surface brightness and
not the flux that determines whether or not a cluster is detected. To
a first approximation, the surface brightness is related to the
cluster flux by the core radius $r_c$, that sets the spatial scale of
the surface brightness distribution. This then enters the XLF through
the computation of the detection volumes of the clusters. Given the
relatively large PSF of the PSPC at the off axis angles considered in
WARPS, we do not expect our results to be sensitive to the choice of
core radius, with the strongest effects expected at the faint end of
the local XLF, which is most sensitive to uncertainties in the
selection function.

For each WARPS cluster, a core radius was estimated from the PSPC data
as the radius at which the surface brightness, fitted by a
$\beta$-model with $\beta=2/3$, is a factor $2^{1/\beta}$ lower than
the central value. The uncertainties on the individual core radii are
large, but the average $r_c$ for WARPS is $\sim100$kpc, whereas
Chandra observations of clusters show an average core radius of $\sim150$
kpc \citep{2008ApJS..174..117M}.

A disadvantage of the Page-Carrera technique is that information about
the core radius of the individual clusters is difficult to
include. Thus when applying this technique, a fixed core radius of
$102$ kpc (average WARPS) was assumed for each cluster. However,
using the $V_\text{max}$ technique it was possible to investigate the
effect of varying $r_c$.  We found that the $V_\text{max}$ technique
yields nearly identical XLFs for a uniform core radius of 100 kpc
(average WARPS), 150 kpc (average Chandra), and the individual core
radii measured from the PSPC data. 

\subsubsection{Uncertainties on the  Local XLF}\label{sec:bayes}
In order to assess the impact of uncertainties on the form of the low
redshift XLF on the measured evolution, we adopted a Bayesian approach
to fitting the XLF. The posterior probability distribution for the set
of model parameters ${\bf \theta}=(\phi^*,L_X^*,\alpha,A,B)$ given the
observed data $\bf{D}$ (the luminosity and redshift of each cluster)
is given in the normal way by
\begin{align}
P({\bf \theta}|\bf{D}) & \propto P(\bf{D}|\bf{\theta})P(\bf{\theta}).
\end{align}
Here the first term on the right is the likelihood function, and the
second term is the prior probability distribution of the model
parameters. This approach allows us to adopt the REFLEX low-z XLF
parameters and their uncertainties as priors on
$\phi^*,L_X^*,\alpha$, which can then be marginalised over. We adopt
weak priors on $A$ and $B$, simply assigning each a Gaussian
distribution with mean zero, and standard deviation 100. We also generalise the
likelihood expression from equation \eqref{eq:likelihood} to include
the statistical scatter of the measured luminosities. This accounts
for the possibility that clusters that are nominally below the flux
limit may be observed to be above the flux limit due to our noisy
measurement of $L_X$ (this is a source of Eddington bias and is
discussed further below).

We divide the $L_X,z$ parameter space into cells $i,j$ with
coordinates ($L_{X,i},z_j$) and widths ($dL_{X,i},dz_j$). As before,
the XLF model predicts a number of clusters in cell $i,j$ as
\begin{align}
  N_{\text{mod},ij} & =
  \lambda(L_{X,i},z_j,{\bf \theta})\mathrm{d}L_{X,i}\mathrm{d}z_j.
\end{align}
However, the final number of clusters expected in cell $i,j$ includes
contributions from all of the other cells in the $L_X$ direction, due
to the noisy measurement of $L_X$. The contribution from a cell at
$L_{X,k},z_j$ to the number counts in a cell at $L_{X,i},z_j$ is
\begin{align}
N_{\text{exp},ijk} & = N_{\text{mod},ik} P(L_{X,j}|L_{X,k},\sigma_k)dL_{X,j}
\end{align}
The probability term here models the measurement noise on a cluster
with ``true'' luminosity $L_{X,k}$ as a Gaussian with a mean $L_{X,k}$
and standard deviation $\sigma_k$. We model the increasing precision
of the luminosity measurement with cluster flux by setting $\sigma_k$
to be inversely proportional to the square root of the flux at
$L_{X,k},z_j$, as expected for measurements dominated by counting
statistics. The constant of proportionality is set to give a $15\%$
luminosity error at the flux limit, in agreement with the observed
clusters.

The final expected number of clusters in cell $L_{X,i},z_j$ is then
\begin{align}\label{eq:nexpij}
N_{\text{exp},ij} & = \sum_k N_{\text{mod},ik} P(L_{X,j}|L_{X,k},\sigma_k)dL_{X,j}
\end{align}
and the likelihood function is then
\begin{align}
P({\bf D}|{\bf \theta}) = \prod_{ij}P(N_{\text{obs},ij}|N_{\text{exp},ij}).
\end{align}
The probability distribution of the number of observed clusters
$N_\text{obs}$ is Poissonian, and can be simplified as before in our
working limit of one or zero observed clusters per cell.

\begin{figure}\begin{center}
\includegraphics[scale=0.42]{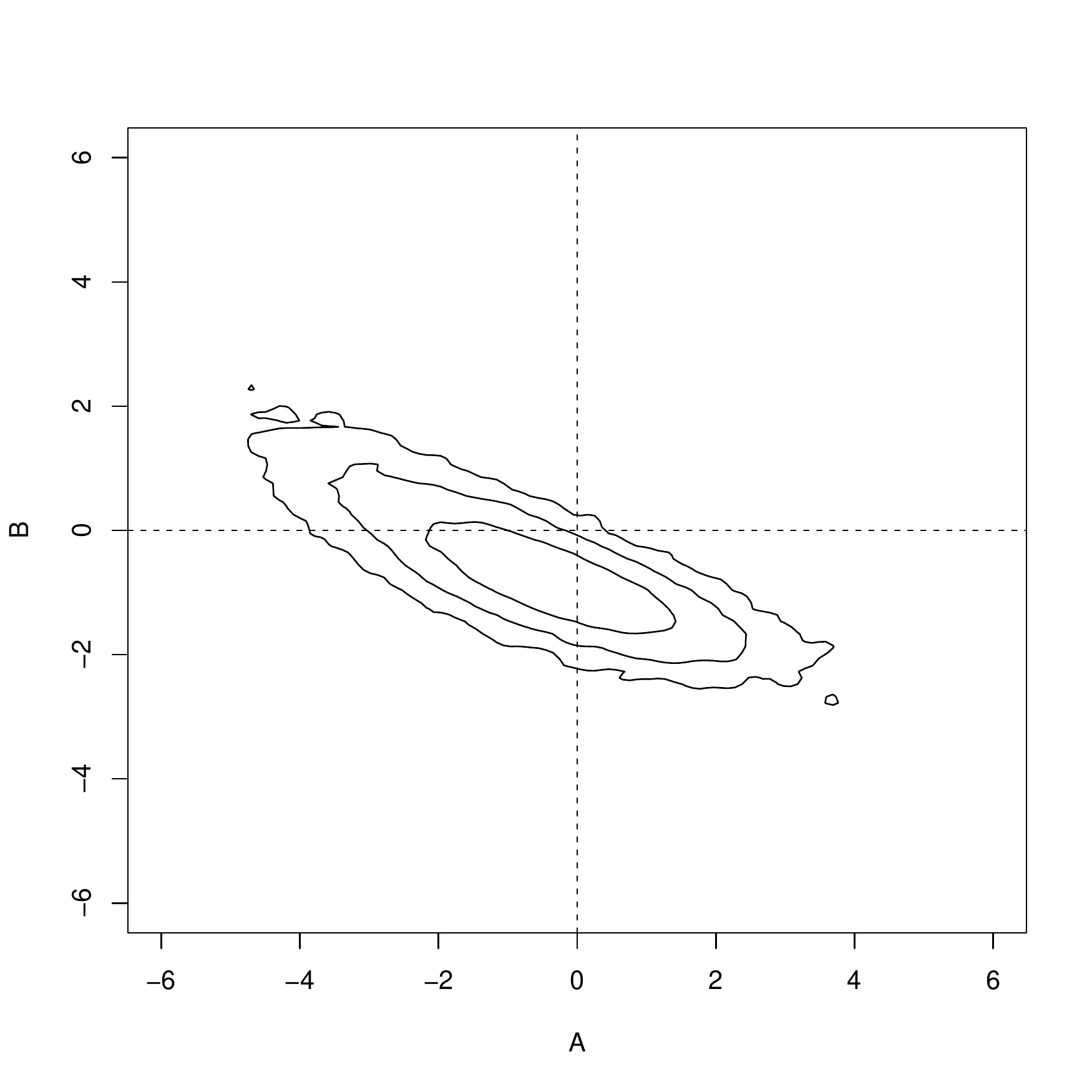}
\caption{Likelihood contours for the evolution parameters $A$ and $B$,
  defined in equations \eqref{eq:evA} and \eqref{eq:evB},
  based on a Bayesian analysis and marginalising over the uncertainty
  on the shape parameters of the local XLF. Light grey contours show
  the constraints from the maximum likelihood analysis. Contours show the
  $1\sigma$, $2\sigma$, and $3\sigma$ confidence limits.}
\label{fig:ABbayes}
\end{center}\end{figure}

\begin{figure}\begin{center}
\includegraphics[scale=0.42]{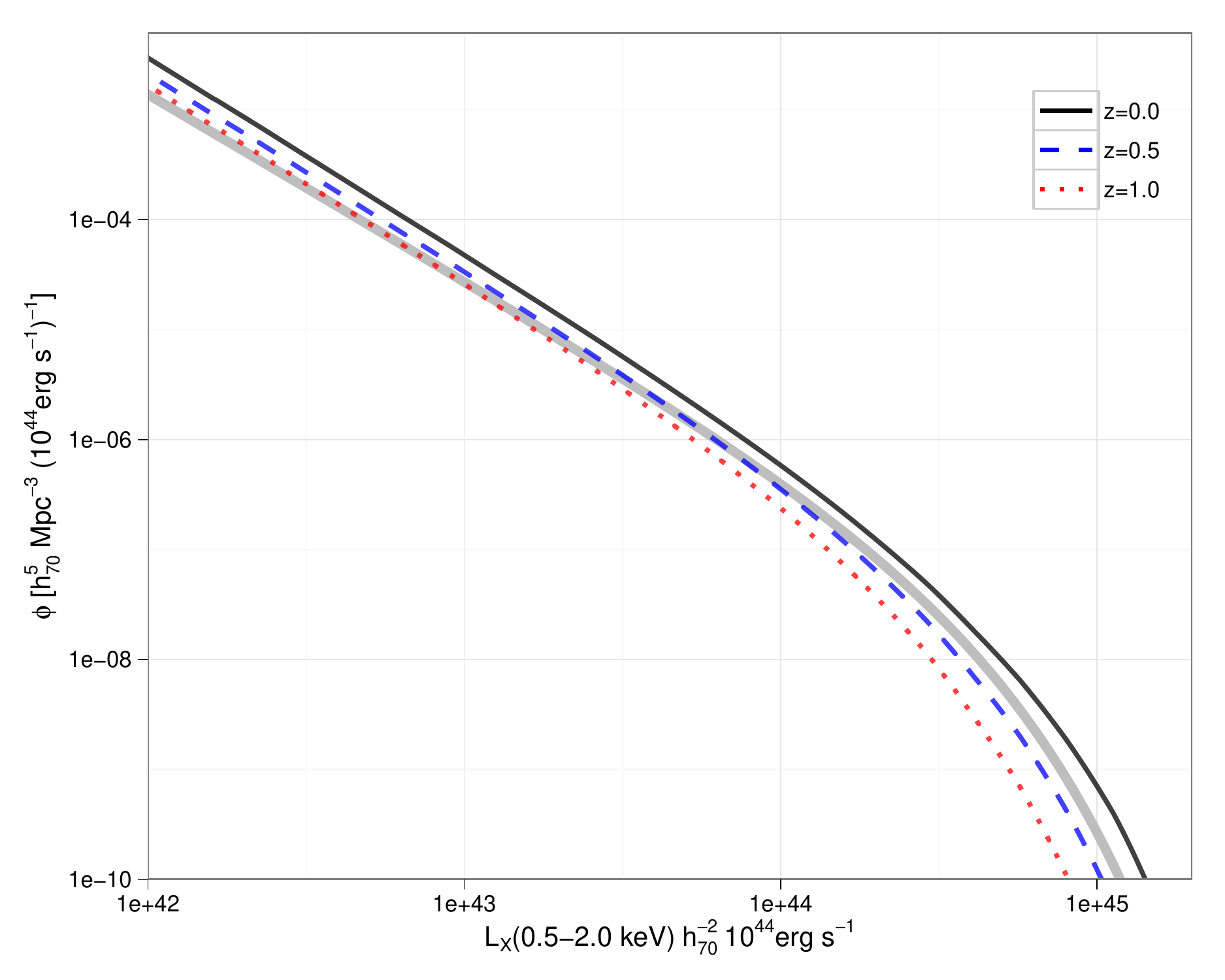}
\caption{Schechter functions with best-fitting evolution parameters
  from the Bayesian analysis. The lines show the form of the Schechter
  function at redshifts 0, 0.5 and 1, and the grey line shows the
  REFLEX Schechter function.}
\label{fig:schecterb}
\end{center}\end{figure}

The posterior probability distribution was analysed using the {\em
  Laplace's Demon} package \cite{hal12} for the {\em R} statistical
computing environment \citep{r12}. An adaptive Metropolis Markov Chain
Monte Carlo algorithm was used, and the resulting constraints on the
model parameters are given in table \ref{tab:xlf}.
The Schechter function shape parameters are all
consistent with the results from the REFLEX data alone, indicating
that the WARPS data do not provide much extra information to constrain
those parameters. The confidence contours for the evolution parameters
are plotted in figure~\ref{fig:ABbayes}. As expected, marginalising
over the uncertainties on the local XLF reduces the precision of the
evolution measurements, though the presence of evolution (i.e. a
difference from $A,B=(0,0)$) is significant at more than $95\%$. This
is the first time that evolution in the cluster XLF has included this
source of uncertainty. The best fitting evolving Schechter function is
compared with the low-redshift REFLEX Schechter function in figure
\ref{fig:schecterb}.

\begin{figure}\begin{center}
\includegraphics[scale=0.42]{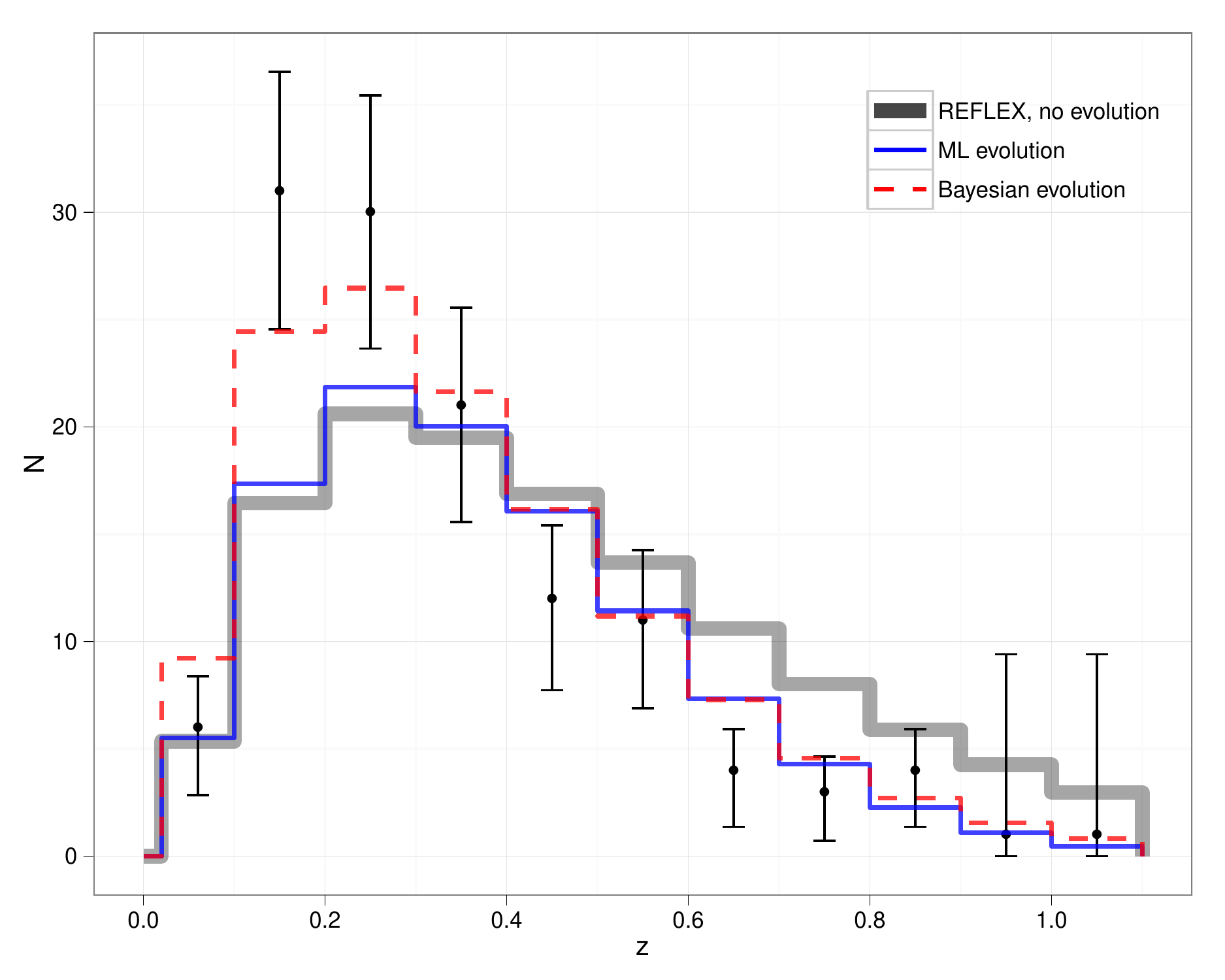}
\caption{The observed redshift distribution of the WARPS clusters is
  compared to the distribution predicted by different models for the
  XLF.  Both the ML and Bayesian fits are to the full sample ($z>0.02$). 
  The error bars on the observed counts 
  are computed according to \citet{geh86}.}
\label{fig:nz}
\end{center}\end{figure}

The different evolution models are plotted in figure~\ref{fig:nz},
which compares the number of clusters as a function of redshift
predicted by the different model XLFs with that observed. The
no-evolution REFLEX model clearly predicts more clusters than observed at
$z>0.6$. There is some tension between the $z>0.3$ maximum
likelihood fit and the Bayesian model, driven by the Bayesian model's
accounting for the excess of WARPS clusters at $0.1<z<0.3$.

\subsubsection{Cluster Correlation Function}
\label{sec:correlationfunction}
We should also consider the likely contamination from any associated
clusters that may lie on the line of sight to the sample
cluster. Since the correlation function for clusters can be written as
$\xi(r) = (r/r_0)^{-1.8}$, with the correlation length $r_0 \sim
20$~Mpc \citep[e.g.][]{Mos01,BasPli04}, we can approximately integrate along the line of sight from,
say, 2~Mpc (to represent the minimum possible separation) to $\sim
150$~Mpc (where the correlation is negligible) to see that we expect
close to twice as many clusters within this range as would be expected
for an unclustered population. (We could alternatively integrate the
two parameter $\xi(r_p, \pi)$ along $r_p =0$ \citep[e.g.][]{Mil99}
to obtain essentially the same result). Taking a column of length $\pm
150$~Mpc and radius 1~Mpc centred on a given cluster (i.e. a volume
$\sim 10^3$~Mpc$^{3}$ and a density of clusters around $10^{-5}
$~Mpc$^{-3}$, appropriate for rather small clusters with only 10-20
bright early type galaxies \citep[see e.g.][]{Koe07}, we
evidently expect only $\sim 1$\% contamination by `clustered
clusters'. Reasonable changes to any of the values used here, will
only change this by a factor of a few. Indeed, if we are interested in
contamination by large clusters (so that masses and fluxes are
seriously affected), the number is around two orders of magnitude
lower still.

\subsection{Excess Number Density in Low-z XLF}
\label{sec:excess}
Figures \ref{fig:XLFlocal}, \ref{fig:Nexpected}, and \ref{fig:nz} show
that the WARPS detects a significant excess of systems in the range
$L_{\text{X}(0.5-2.0\text{keV})} = 1.0 - 2.0 \times 10^{43}$ erg
s$^{-1}$ and $0.1<z<0.3$ relative to the REFLEX Schechter
function. These luminosities correspond to $\sim2$keV systems, so are
poor clusters or galaxy groups. Interestingly, a significant excess is
seen at the same luminosity range in the 160SD low-z XLF. In
determining this excess, we have been comparing observed number counts
to the best fitting model to the observed REFLEX XLF. It is worth
considering whether an excess is present in the REFLEX data, but
inspection of figure 4 in \citep{mul04} shows that the REFLEX data are close to the
best fitting Schechter function in this luminosity range. 
(note their figure is for $H_0=50$ km s$^{-1}$ Mpc$^{-1}$).

\begin{figure}\begin{center}
\includegraphics[scale=0.42]{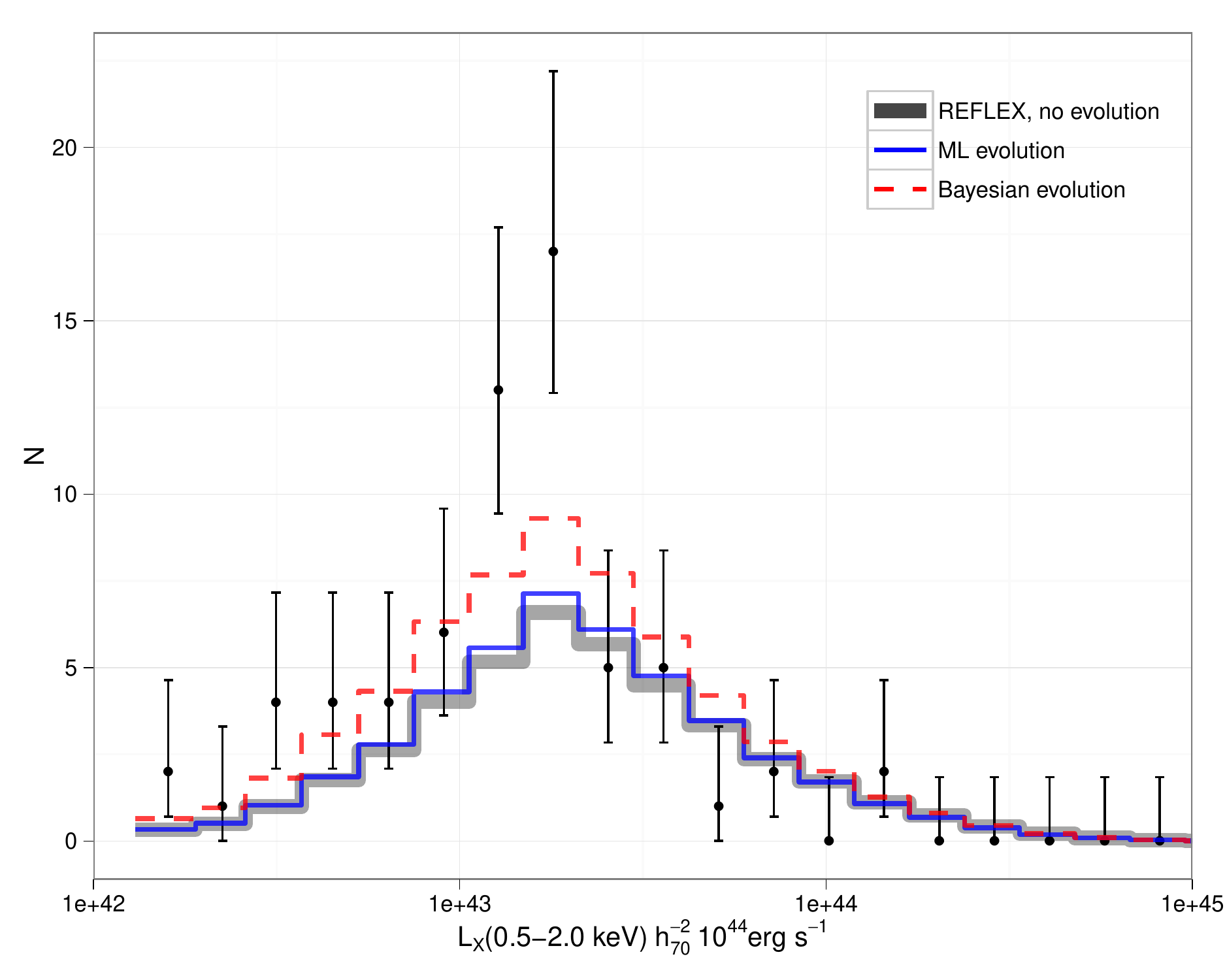}
\caption{The observed luminosity distribution of the WARPS clusters at
  $z<0.3$ is compared to the distribution predicted by different
  models for the XLF. The error bars on the observed counts are
  computed according to \citet{geh86}.}
\label{fig:nL2}
\end{center}\end{figure}

The best-fitting Bayesian XLF model predicts a significantly larger
number of clusters at $z<0.3$ than the REFLEX model, as shown in
figure~\ref{fig:nL2}. This somewhat reduces the significance of the
excess clusters in the bump at $L_{\text{X}(0.5-2.0\text{keV})} =
1.0 - 2.0 \times 10^{43}$ erg s$^{-1}$. Marginalised over the
uncertainties on the model parameters, the Bayesian model predicts
$16.5\pm3.7$ clusters, compared with the 28 observed. The probability
of observing at least 28 clusters in this luminosity bump of
$2\%$. There is thus suggestive evidence that a real excess remains,
and we now consider possible factors that could contribute to this.

Could the excess be simply a result of sample (cosmic) variance? The survey
areas of WARPS and 160SD are much smaller than REFLEX, but the volumes
surveyed are still significant. The clusters in the bump feature
occupy the redshift range $0.1<z<0.3$ and the volume surveyed by WARPS
in this redshift range is $\sim10^7h_{70}^3$ Mpc$^{3}$. This volume is larger 
than the volumes in which sample variance is
expected to be significant for galaxy surveys (Driver \& Robotham
2010). For cluster surveys,  \cite{HuKrav03} provide analytical
approximations to compute the relative contributions of Poisson noise
and sample variance. Approximating the WARPS as a volume-limited
survey with a mass threshold of $10^{14}M_\odot$ at $z=0.2$, the effects
of Poisson noise and sample variance are approximately equal, with
Poisson noise dominating at higher redshifts, and sample variance
dominating at lower redshifts. This indicates that sample variance
could be responsible for the excess cluster counts in the
WARPS. However, it is more difficult for sample variance to explain
the coincident, stronger, excess seen in the same part of the $L,z$
plane in the 160SD survey. Only 40$\%$ of the WARPS fields and $35\%$ of the
WARPS clusters are in common with 160SD. If the excess were due to
sample variance, then the addition of extra, independent, fields should
result in regression to the mean, not increased significance of the
excess as observed. We thus conclude it is unlikely that the excess
seen in both surveys is due to sample variance, and investigate other
possibilities.

The three dimensional distribution of the clusters in the luminosity
and redshift range of the excess were examined, but there was no
evidence for clustering in volume, so the excess is not caused by a
superstructure of clusters. 
This is expected, since the selected fields are scattered across a large fraction of the sky.

A further possibility to explain the excess numbers is contamination in the detected flux from unresolved X-ray point sources, for example low luminosity Active Galactic Nuclei (AGN). Such contamination was removed where possible, but not all contaminating sources are resolved in
the ROSAT PSPC images. Hence, some residual contamination is expected, enhancing the estimated cluster luminosities. Detailed modelling of the AGN population is beyond the scope of this paper, but this contamination would differ from the scatter models discussed above, as the effect is purely additive. Some mass or redshift dependence of the AGN contamination may be required to manifest the localised excess of clusters in the $L_{X},z$ plane.

Finally, we consider if selection bias may be responsible for the
excess clusters seen in this region of the $L_X,z$ plane. This is
plausible, given that the excess is close to the flux limit in the
region of the $L_X,z$ plane where the WARPS is most sensitive (see
figure \ref{fig:sample}). Eddington
bias enhances cluster number counts when scatter is present in the
luminosities of the population. The slope of the XLF means that for a
given flux limit, there are more clusters below the flux limit that
may scatter into the sample than above the flux limit that may scatter
out of the sample. There are two sources of scatter that may be important:
statistical scatter due to the counting statistics on the $L_X$
measurement, and intrinsic scatter in the cluster population. Our
Bayesian analysis allows us to investigate each of these sources of
scatter in turn, by modifying the model for the population scatter in
equation \eqref{eq:nexpij}.

Note that if the population scatter is constant with $L_X$ and $z$,
then a bias is present at all redshifts and luminosities, and
increases towards higher luminosities due to the steepening of the
XLF, so would not produce a localised excess in the $L_x,z$
plane. However, the statistical scatter decreases with increasing
$L_X$, as $\sim\sqrt{N}$ errors decrease above the flux limit, and
increase below it. This gives rise to a bias that is strongest near
the flux limit, with counting statistics allowing clusters nominally
too faint for the sample to appear above the flux limit, so could
plausibly contribute to the observed excess. However, the typical
measurement error on luminosities close to the flux limit is
$\approx15\%$, and even modelling the increasing flux scatter below the
flux limit, the bias due to this source of scatter was found to
contribute $<1$ additional cluster in the regions of the
excess, compared to models with no scatter. We note however, that our
model for the scaling of the statistical scatter with flux is
simplistic -- the measurement errors also depend on the exposure time
and background level in the source field. Modelling this for clusters
below the flux limit would require extensive simulations and is beyond
the scope of this work.

The intrinsic scatter in luminosity of the cluster population is known
to be significant, and if this varies with mass or redshift, it could
result in a bias that contributed to a localised excess in number
counts. The variation of cluster scatter with mass is not well
measured, but there is evidence that the intrinsic luminosity scatter
decreases above $z\approx0.4$ \citep{mau07b}, albeit measured in a
heterogeneous sample. We test the effect that evolving scatter could
have on the measured XLF by replacing the scatter model in equation
\eqref{eq:nexpij} with a lognormal distribution, with a standard
deviation of $50\%$ at $z<0.45$, decreasing smoothly to $20\%$ at
$z>0.55$ \citep{mau07b}. The effect of this evolving scatter is to
increase the number of $z<0.3$ clusters predicted by the Bayesian
model by $\approx10\%$, but this still leaves a significant excess of
observed clusters at around $L_{\text{X}(0.5-2.0\text{keV})} = 1.0 - 2.0
\times 10^{43}$ erg s$^{-1}$.

Similarly, introducing an ad hoc model of intrinsic scatter which
decreases from $90\%$ to $20\%$ at $L_{\text{X}(0.5-2.0\text{keV})} = 2.5
\times 10^{43}$ erg s$^{-1}$ (designed to maximise the Eddington bias
effect), only serves to increase the model prediction by $2.5$
clusters in the bins contributing to the excess. When compared to the
REFLEX XLF, the observed excess of WARPS clusters remains highly
significant in the face of all bias contributions.

A luminosity of $2\times 10^{43}$ erg s$^{-1}$ corresponds to a mass of $M_{200} \simeq 10^{14} M_\odot$, at
the borderline between groups and clusters. It is possible that
some feature of the cluster population in this region of the $L_\text{X},z$ plane
enhances their detectability, by e.g. enhancing their surface
brightness. This could result from, for instance, an enhanced AGN or cool core
population, but the effects would need to be relatively localised
in $L_\text{X}$ and z. With the present data it is not possible to
investigate this further, but with the arrival of new deep cluster
surveys such as XXL or XCS, it should be possible to verify these
results and extend the investigation.

\subsection{Sensitivity to High-Redshift Cool Core Clusters}
Since the detection of clusters in X-ray surveys is driven by their
surface brightness, the presence or absence of centrally peaked
emission associated with a cool core may have a strong impact on the
detection of a given cluster. There is some debate in the literature
as to whether there are significant numbers of cool core clusters in
the high-z ($z>0.5$) Universe. Both \citet{vik06c} and
\citet{san10} find a lack of high-z cool core clusters in the distant
400SD sample, but \citet{san10} find evidence for moderate cool cores
in high-z clusters detected in the WARPS and RDCS. They argue that the
400SD (and by extension 160SD) may have discarded high-z cool core
clusters as being unresolved, but note that this does not imply
incompleteness in the 400SD providing the surface brightness
dependence was modelled into the selection function.

The excellent agreement of the evolution in the XLF seen in the WARPS
and 160SD at $z>0.3$ despite their very different cluster detection
algorithms implies that this is indeed the case, and that there are no
significant problems with the selection function of either survey. We
also note that the comparisons of clusters detected or missed by WARPS
and 160SD in 157 common fields revealed no significant discrepancies,
with differences being explained by the differing selection criteria
\citep{hor08}.

\section{Conclusions}

We measured the evolution of the XLF out to $z\sim1$ from the combined
WARPS-I and WARPS-II surveys, finding significant evidence for
negative evolution, in the sense of a reduction in the number density
of massive luminous clusters with redshift.  This is confirmed by
comparing expected and observed numbers, and more convincingly by the
maximum likelihood analysis. This is consistent with previous
measurements of the evolution of the XLF, and the expectations of
hierarchical structure formation in a $\Lambda$CDM Universe.

We investigate the sensitivity of these results to various sources of
systematic uncertainty affecting the WARPS XLF and selection
function. The results are not significantly affected by the modelling
of the core radii of the clusters, the assumed upper redshift limit of
the survey, or the technique used for estimating detection volumes for
the clusters. The assumed value of $z_\text{max}=1.1$ is fairly
conservative, as there is a cluster detected at $z=1.05$ and many of
the clusters would be detectable in X-rays to significantly higher
redshift. Thus the true evolution could be somewhat stronger than we
measure.

For the first time, we fully incorporate the uncertainties on the
low-redshift XLF into a Bayesian analysis of the evolution, and find
that while the precision of the measurements is reduced, evolution is
still significant at the $95\%$ level.

The good agreement of the measured evolution in the WARPS, 160SD and
other surveys suggest that the result is not sensitive to the details
of the cluster detection and follow-up strategy, and that the
selection functions of both surveys are accurately modelled, including
the effects of cool cores on the detectability of clusters.

We identified a significant excess of $\sim2$keV systems at $z<0.3$
observed in both the WARPS and 160SD surveys relative to the REFLEX
XLF. A Bayesian fit to the WARPS data, which uses the REFLEX
measurements as priors yields a model with slightly higher $\phi^*$
and $\alpha$ values than REFLEX, and reduces the excess. However, even
with this model, and with including possible contributions from
Eddington bias, the excess remains significant. The cause of the
excess in not clear at present, although its presence in both the
WARPS and 160SD argue against it being a result of a mis-calibrated
selection function. New, more sensitive measurements of the XLF with
surveys like $XXL$ \citep{2011MNRAS.414.1732P} and $XCS$
\citep[][]{2011arXiv1106.3056M} will provide better statistics for
this part of the $L_X,z$ plane, providing a means to address this
question further.

\section*{Acknowledgments} We thank Chris Collins and Andy Young for
useful discussions. LK acknowledges support from the European Research
Council under the EC FP7 grant number 240185.

\appendix
\section{XLF Tables}
For future reference we present tables of the binned X-ray Luminosity Function. Per bin we also quote the number of observed clusters, their median redshift $\tilde{z}$, and their average X-ray luminosity $\overline{L}_\text{X}$, see Table \ref{tab:xlf}. 

\begin{table*}
\renewcommand{\arraystretch}{1.2} 
\caption{The X-ray Luminosity Function for the local, the intermediate redshift, and high redshift universe as measured by WARPS. Also shown are the observed number of clusters $N_\text{obs}$, their median redshift $\tilde{z}$, and their average luminosity $\overline{L}_\text{X}$. 
}
\centering
\begin{tabular}{lllllll}\\ \hline
$L_\text{X,centre}$             & $\phi$                                        & $\phi^{+1\sigma}$                 & $\phi^{-1\sigma}$                  & {}                 & & $\overline{L}_\text{X}$ \\
(0.5-2.0 keV)                        & [$h_{70}^5 \text{Mpc}^{-3}$  & [$h_{70}^5 \text{Mpc}^{-3}$ & [$h_{70}^5 \text{Mpc}^{-3}$ & {}                 & & (0.5-2.0 keV) \\
$[h_{70}^{-2}$ 10$^{44}$  & $(10^{44} \text{erg}$              & $(10^{44} \text{erg}$             & $(10^{44} \text{erg}$             & {}                 & & $[h_{70}^{-2}$ 10$^{44}$ \\
erg s$^{-1}$]                         &  s$^{-1})^{-1}]$                        &  s$^{-1})^{-1}]$                      &  s$^{-1})^{-1}]$                        & $N_\text{obs}$ & $\tilde{z}$ & erg s$^{-1}]$  \\
 \hline\hline
\\
$0.02<z<0.3$ \\
\\ 
0.011 & 2.26 $\times$ $10^{-3}$ & 7.46 $\times$ $10^{-2}$ & 3.91 $\times$ $10^{-4}$ & 1 & 0.051 & 0.010 \\
0.016 & 1.14 $\times$ $10^{-3}$ & 2.64 $\times$ $10^{-2}$ & 4.04 $\times$ $10^{-4}$ & 2 & 0.063 & 0.015 \\
0.022 & 2.33 $\times$ $10^{-4}$ & 7.69 $\times$ $10^{-4}$ & 4.03 $\times$ $10^{-3}$ & 1 & 0.107 & 0.025 \\
0.031 & 3.46 $\times$ $10^{-4}$ & 6.20 $\times$ $10^{-4}$ & 1.80 $\times$ $10^{-4}$ & 4 & 0.125 & 0.032 \\
0.044 & 1.33 $\times$ $10^{-4}$ & 2.38 $\times$ $10^{-4}$ & 6.94 $\times$ $10^{-5}$ & 4 & 0.119 & 0.043 \\
0.063 & 5.72 $\times$ $10^{-5}$ & 1.02 $\times$ $10^{-4}$ & 2.98 $\times$ $10^{-5}$ & 4 & 0.151 & 0.060 \\
0.089 & 3.53 $\times$ $10^{-5}$ & 5.64 $\times$ $10^{-5}$ & 2.13 $\times$ $10^{-5}$ & 6 & 0.150	& 0.090 \\
0.125 & 3.68 $\times$ $10^{-5}$ & 5.01 $\times$ $10^{-5}$ & 2.67 $\times$ $10^{-5}$ & 13 & 0.193 & 0.125 \\
0.177 & 2.25 $\times$ $10^{-5}$ & 2.94 $\times$ $10^{-5}$ & 1.71 $\times$ $10^{-5}$ & 17 & 0.226 & 0.174 \\
0.251 & 4.85 $\times$ $10^{-6}$ & 8.13 $\times$ $10^{-6}$ & 2.75 $\times$ $10^{-6}$ & 5 & 0.252 & 0.233 \\
0.355 & 3.24 $\times$ $10^{-6}$ & 5.43 $\times$ $10^{-6}$ & 1.84 $\times$ $10^{-6}$ & 5 & 0.240	& 0.352 \\
0.502 & 4.82 $\times$ $10^{-7}$ & 1.59 $\times$ $10^{-6}$ & 8.34 $\times$ $10^{-8}$  & 1 & 0.242	& 0.457 \\
0.710 & 6.51 $\times$ $10^{-7}$ & 1.51 $\times$ $10^{-6}$ & 2.30 $\times$ $10^{-7}$  & 2 & 0.244	& 0.741 \\
1.004 & 0	& 8.98 $\times$ $10^{-7}$ & 0 & 0 & $n.a.$ & $n.a.$ \\
1.420 & 3.25 $\times$ $10^{-7}$ & 7.54 $\times$ $10^{-7}$ & 1.15 $\times$ $10^{-7}$  & 2 & 0.292 & 1.351 \\
\\ \hline
\\
$0.3<z<0.6$ \\
\\
0.177 & 2.07 $\times$ $10^{-5}$  &	4.79 $\times$ $10^{-5}$  & 7.31 $\times$ $10^{-6}$  & 2 & 0.304 & 0.196 \\
0.251 & 4.23 $\times$ $10^{-6}$  &	8.35 $\times$ $10^{-6}$  & 1.93 $\times$ $10^{-6}$ 	 & 3 & 0.315 & 0.259 \\
0.355 & 2.78 $\times$ $10^{-6}$  &	4.28 $\times$ $10^{-6}$  & 1.75 $\times$ $10^{-6}$ 	 & 7 & 0.370 &	0.373 \\
0.502 & 1.31 $\times$ $10^{-6}$  &	1.96 $\times$ $10^{-6}$  & 8.56 $\times$ $10^{-7}$  & 8 & 0.378 & 0.507 \\ 
0.710 & 6.68 $\times$ $10^{-7}$  &	9.53 $\times$ $10^{-7}$  & 4.60 $\times$ $10^{-7}$ 	 & 10 & 0.461 & 0.707 \\
1.004 & 3.54 $\times$ $10^{-7}$  &	5.29 $\times$ $10^{-7}$  & 2.32 $\times$ $10^{-7}$ 	 & 8 & 0.502 & 1.006 \\
1.420 & 8.94 $\times$ $10^{-8}$  &	1.76 $\times$ $10^{-7}$  & 4.08$\times$ $10^{-8}$ 	 &  3	& 0.500 &	1.455 \\
2.008 & 4.42 $\times$ $10^{-8}$  &	1.02 $\times$ $10^{-7}$  & 1.56$\times$ $10^{-8}$ 	 & 2	& 0.561 &	1.909 \\
2.840 & 1.56 $\times$ $10^{-8}$  &	5.16 $\times$ $10^{-8}$  & 2.71 $\times$ $10^{-9}$ 	 & 1	& 0.517 &	2.730 \\
\\ \hline
\\
$0.6<z<1.1$ \\
\\
1.004 & 1.01 $\times$ $10^{-7}$  & 3.33 $\times$ $10^{-7}$  & 1.74 $\times$ $10^{-8}$  & 1 & 0.679 & 1.110 \\
1.420 & 7.34 $\times$ $10^{-8}$ 	& 1.45 $\times$ $10^{-7}$  & 3.35 $\times$ $10^{-8}$  & 3 & 0.722 & 1.421 \\
2.008 & 4.87 $\times$ $10^{-8}$ 	& 8.17 $\times$ $10^{-8}$  & 2.77 $\times$ $10^{-8}$  & 5 & 0.832 & 2.114 \\
2.840 & 0	& 2.51 $\times$ $10^{-8}$  & 0 & 0 & $n.a.$ & $n.a.$ \\
4.016 & 5.88 $\times$ $10^{-9}$ 	& 1.36 $\times$ $10^{-8}$  & 2.08$\times$ $10^{-9}$  & 2 & 0.820 & 4.164 \\	
5.680 & 2.08 $\times$ $10^{-9}$ 	& 6.86 $\times$ $10^{-9}$  & 3.60 $\times$ $10^{-10}$  & 1 & 0.833 & 6.655 \\
8.032 & 1.47 $\times$ $10^{-9}$ 	& 4.85 $\times$ $10^{-9}$  & 2.54 $\times$ $10^{-10}$  & 1 & 0.892 & 9.271 \\
\end{tabular}

\label{tab:xlf}
\end{table*}
\bibliographystyle{mn2e}
\bibliography{clusters.bbl}

\bsp

\label{lastpage}

\end{document}